\newif\ifOneCol

\ifOneCol
\documentclass[onecolumn,11pt]{IEEEtran}
\else
\documentclass[twocolumn,10pt]{IEEEtran}
\fi 

\usepackage[utf8]{inputenc}
\usepackage{graphicx}
\usepackage{amsmath}

\usepackage{amssymb}
\usepackage[version=4]{mhchem}
\usepackage{siunitx}
\usepackage{longtable,tabularx}
\usepackage{soul}
\usepackage{color}
\usepackage{amsthm}
\usepackage[english]{babel}
\newtheorem{theorem}{Theorem}[section]

\newtheorem{remark}[theorem]{Remark}

\begin{document}


\title{UAV Optimal Guidance in Wind Fields Using ZEM/ZEV with Generalized Performance Index}
\author{Ely C. de Paiva, Bruno Carvalho, Luis Rodrigues
\thanks{E. C. Paiva is with the School of Mechanical Engineering, University of Campinas.\textit{Address: DSI, FEM-Unicamp, R. Mendeleyev, 200, Campinas, SP, Brazil, CEP 13083-860, email:{ely.paiva@fem.unicamp.br}}}
\thanks{B. Carvalho and L. Rodrigues are with the Department of Electrical and Computer Engineering, Concordia University. \textit{Address: 1515 St. Catherine W., Montréal, QC, Canada H3G 2W1, email:{b\_per@encs.concordia.ca, luis.rodrigues@concordia.ca.}}}
}

\maketitle

\begin{abstract}
\textcolor{black}{This paper presents an optimal guidance approach for a UAV navigation between two given points in 3D considering the wind influence. The proposed cost function to be minimized involves the weighting of the travel time and the control energy.  
An analytical expression is derived for the optimal cost yielding a fourth order polynomial whose positive real roots correspond to the optimal travel times.
The optimization problem is shown to be equivalent to the  Zero-Effort-Miss/Zero-Effort-Velocity  (ZEM/ZEV) optimal guidance approach for the case of a constant wind acceleration. Case studies for the rendez-vous and the intercept problems are shown through simulation examples for different wind conditions.}
\end{abstract}


\section{Introduction}
\label{sec:introduction}

 In the last decade there has been a growing interest in the development of Flight Management Systems (FMS) for unmanned aerial vehicles (UAVs), with special emphasis on task and path planning, as well as optimal guidance \cite{Kumar2,2017SurvayUAV,2016Villa}.
In a top-down layer view, a UAV flight management system usually consists of the mission planning, the path planning and the guidance system \cite{tsourdos2010}. Upon receiving a set of waypoints to be reached, the automatic guidance system provides the acceleration commands to the low level flight controller in order to minimize a predefined performance criteria such as, for instance, the traveled length, the traveled time, or the consumed energy \cite{Zarchan2007, 2011BookGuidance}.  Additionally, to cope with the need of path replanning, it is desirable to have a computationally efficient FMS in the onboard real-time embedded system  \cite{tsourdos2010}.
Therefore, a closed-form solution for the  guidance algorithm is sometimes preferred instead of a numerical one. 


In this paper, we consider the design of a time-energy optimal trade-off guidance for a vehicle that flies between two given waypoints in 3D under time-varying wind fields. We consider that the motion of the vehicle is described by a point-mass linear kinematic model, which is a common assumption when the trajectories are to be computed as reference signals for the low level flight controller \cite{2011BookGuidance}.  The idea here is to evolve classical missile guidance concepts to design motion planning. Most of the previous approaches, with this purpose,  assume a constant vehicle velocity and/or are based on non-optimal guidance laws \cite{2005WaypointUAV, 2017ChinaWaypoint,Fossen2007, 2016WaypointTexas, 2011WaypointSlovakia}.
 


\textcolor{black}{A classical guidance technique is called Proportional Navigation (PN), originally designed for missile interception in the 1940's \cite{Zarchan2007}.
The main idea of PN is to generate an orthogonal velocity command proportional to the rate of change of the Line-of-Sight  angle (LOS) between target and pursuer. Indeed, besides the many non-optimal PN variants, one can also find some types of optimal PN, for example, when a performance index weighting the squared orthogonal acceleration and the lateral miss error (ZEM distance) is optimized \cite{2011BookGuidance,2019TAES1,TAES2010}.
The ZEM error is defined as the final attained position error between pursuer and target if there is no corrective control acceleration \cite{TAES2018}.}


\textcolor{black}{
The optimal guidance approach proposed in this paper is a generalization of the classical Linear Quadratic Minimum Time problem (LQMT) for the case of relative velocities. The development of LQMT comes from the decade of 1960, and it was the basis of one of the moon landing guidance approaches, known as E-guidance \cite{2019PingLu}, aiming at minimum-fuel consumption.  In 1997, D'Souza \cite{DSouza1997} investigated the LQMT problem for the planetary landing (rendezvous), deriving a polynomial equation for the optimal time-to-go, which is extended in our work for the relative velocities case. 
In 2008, a new optimal guidance law was introduced by Ebrahimi et al. \cite{Ebrahimi2008}, through the definition of the zero-effort-velocity (ZEV)
which is analogous to the the ZEM distance error, as it corresponds to the velocity error (pursuer-target) at the end of a given mission if no further control acceleration is provided. 
The ZEM/ZEV approach of Ebrahimi is an extension of the work of D'Souza, as it is able to treat the case of a non-uniform gravity model \cite{Guo2012Generalz,Zheng2017Review,Furfaro2}. In our case,  as conceptually identical, we use the formulation of ZEM/ZEV  subtituting the gravity for the wind acceleration in the vehicle model. }



In fact, ZEM/ZEV appears more frequently in the context of space applications, as the simplified analytical model of the pursuer  usually does not consider the effect of the atmospheric drag, although this can still be included numerically \cite{Furfaro2017,2016Kepler}. 
However, the wind effect is an important nonlinear disturbance for a UAV, especially for the smaller vehicles, such as an autonomous blimp,  which motivated the current work \cite{de2006project,moutinho2016airship}.
\textcolor{black}{If on one hand, the wind can affect the UAV maneuvering capabilities (such as the minimum turning radius), on the other hand, it may also help to save energy  \cite{2017Lanteigne} by flying in the wind direction. 
A number of researchers have investigated the problem of guidance in the presence of wind \cite{2017SurvayUAV,2005WaypointUAV,2005Dubin,2017DelftWind,2009AirshipStratos,2016RendezvousNum,2007Dubin,2013BakolasDubin}. However, most articles in the open literature are focused on minimum-time problems, usually under constant airspeed, and an analytical "time-energy" optimal solution for the rendez-vous/intercept problem is hard to find.}

A pioneering work with variable airspeed was proposed in the paper of Bakolas \cite{Bakolas2014} where the authors address the minimum-time intercept guidance of an isotropic rocket in the presence of wind and subject to a norm-constrained acceleration. For the general wind case, however, the solution is obtained numerically using intense computational effort, and it is strongly dependent on the initial point.  Moreover, the approach is limited to the minimum-time case, and the final velocity is supposed to be free (intercept) instead of constrained (rendez-vous case).
\textcolor{black}{Our proposed methodology removes this constraint  allowing to have both the position and the velocity constrained at the target waypoint, assuming a linear time-variant wind. For the general wind model case, a sub-optimal solution can be found using an iterative procedure, considering piecewise linear time-varying wind speed regions}.

The contributions of this paper are:
\begin{enumerate}
\item the solution of the optimal control problem for both rendez-vous and intercept maneuvers, under the effect of wind, using a generalized performance functional that trades off the norm of the acceleration and the flight time \textcolor{black}{(Sec.III).} 
\item analytical expressions for both the optimal control input and the optimal cost \textcolor{black}{(Sec.III,IV).}
\item \textcolor{black}{ the generalization of the LQMT problem  \cite{Lewis1991},\cite{DSouza1997} to the case of relative velocities  (Sec.IV).} 
\item  \textcolor{black}{an iterative proposal for the general wind model, approached by piecewise linear time-varying wind speeds.}
\item the generalization of the optimal feedback guidance approach of ZEM/ZEV to the relative velocity case under constant wind accelerations \textcolor{black}{(Sec.V).}
\item a detailed discussion providing insight on the  existence of single, multiple, or no solutions for the optimal flight time for the case of minimum energy
\textcolor{black}{(Sec.VI), complementing results from 
\cite{Guo2012Generalz,Hawking2013,Guo2019}.}
\end{enumerate}


\textcolor{black}{
The rest of the paper is organized as follows. In Section II, we introduce
the problem formulation. Sections III presents the general solution of the optimization problem. Section IV details the solution for some  wind cases. Section V shows the optimal solutions using ZEM/ZEV. Section VI provides a discussion about the existence of solutions. Section VII shows simulation results, and section VIII presents the conclusions.
}


\section{Problem Formulation}
\label{two}

\begin{figure*}[hbt]
\noindent \centering{}
\includegraphics[scale=0.32]{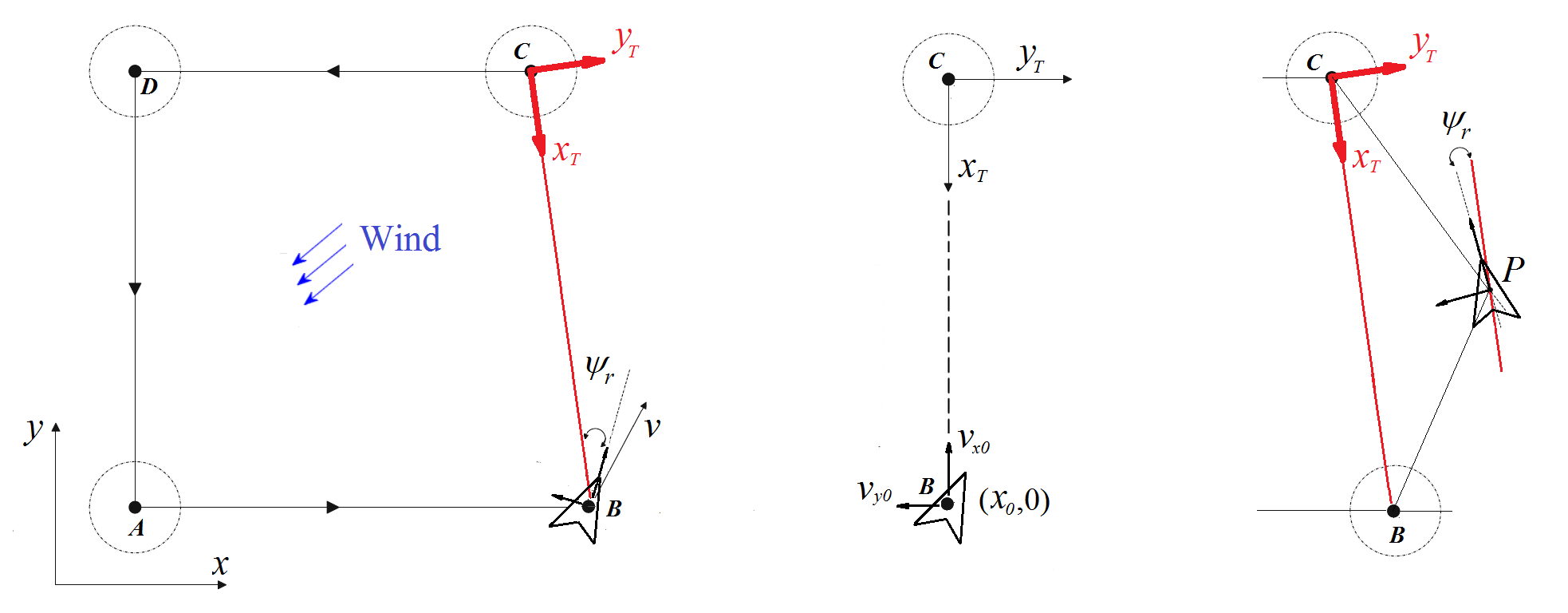}\caption{Problem of a general waypoint navigation in wind fields, with emphasis on the optimal path between two given points.  }
\label{Figure2}
\end{figure*}

We present here the problem statement related to
the optimal 3D trajectory generation for an aerial vehicle (UAV), under the wind influence.
When travelling between two given points, for example from point B to point C (Figure 1), \textcolor{black}{illustrated here in 2D for simplification},  we consider
the position and velocity of the UAV as referenced in the inertial target waypoint $x_T-y_T$. 
Without loss of generality we define such a frame so that we can assume that the initial lateral coordinates are zero (middle picture of fig. \ref{Figure2}).
The objective is to optimize a given cost function, weighting time/energy, while  travelling between these two points. The UAV can arrive at the target point at a constrained terminal velocity (rendez-vous problem),
or at a free terminal velocity (intercept problem). 


Let the motion of the UAV be described by the equations:
\begin{equation}
\begin{array}{l}
{{\dot{\bf  r}}}(t)={\bf {v}}_{g}(t)\\
{\dot{\bf  v}}_{g}(t)={{\dot{ \bf v}}}_{a}(t)+{ \dot{\bf {w}}}(t)\\
{\bf {u}}(t)={{\dot{\bf  v}}}_{a}(t)
\end{array}\label{eq:model}
\end{equation}
where vector ${\bf {r}}(t)=[x(t),\,\,y(t),\,\,z(t)]^{T}\in\ensuremath{\mathbb{R}^{3}}$
is the UAV position in the given inertial frame, ${\bf {v}}_{g}(t)=[v_{gx}(t),v_{gy}(t),v_{gz}(t)]^{T}\in\ensuremath{\mathbb{R}^{3}}$ is the inertial velocity vector,
${\bf {v}}_{a}(t)=[v_{ax}(t),v_{ay}(t),v_{az}(t)]^{T}\in\ensuremath{\mathbb{R}^{3}}$
is the velocity relative to the air (airspeed) ${\bf {w}}(t)=[w_{x}(t),\,\,w{}_{y}(t),\,\,w{}_{z}(t)]^{T}\in\ensuremath{\mathbb{R}^{3}}$
is the wind velocity vector, and ${\bf {u}}(t)$ is the control input
corresponding to the acceleration relative to the air. \\ 

\noindent
\begin{remark}\label{rem:generality}
Note that the model (\ref{eq:model}) is quite general and can also represent many other guidance and control scenarios. For example it can represent orbital rendez-vous and intercept missions where ${\bf u}$ is the acceleration due to thrust and $\dot {\bf w}$ is the gravitational acceleration (see \cite{Guo2012Generalz}). Additionally, the same model can represent the transport of fragile packages with ${\bf u}$ being the acceleration caused by aerodynamic forces (which should be minimized to avoid damage of the fragilie package) and $\dot {\bf w}$ the acceleration due to gravity (see \cite{YuanRodrigues2019}).
\end{remark}
The optimal control problem for the trajectory generation between
two waypoints minimizing a performance index that trades off control energy and flight time can be formulated as: 
\begin{equation}
\begin{array}{rl}
\min_{{\bf u},t_{f}}\,&\int_{0}^{t_{f}}\left(\frac{1}{2}{\bf u}^{T}{\bf u}+C_{I}\right)d\tau\\
s.\,t.&(\ref{eq:model})\\
& {\bf r}(0)={\bf r}_{0},~~ {\bf r}(t_{f})={\bf r}_{f} \\
&{\bf {v}}_{g}(0)={\bf v}_{g0}\\
& {\bf v}_{g}(t_{f})={\bf v}_{gf}~~\mbox{(rendez-vous case)}\\
&  {\bf v}_{g}(t_{f})={\rm free}~~\mbox{(intercept case)}\\
&{\bf w}(0)={\bf w}_0,~~{\bf w}(t_f)={\bf w}_f\\
\end{array}\label{eq:statement}
\end{equation}
where $C_{I}>0$ is the trade-off coefficient between the costs associated
to  ${\bf u}^{T}{\bf u}$ and the
 flight time $t_{f}$.
Using a higher $C_{I}$ one puts more weight on the total flight time.
Conversely,  with a lower $C_{I}$ one puts more weight on the  control energy. \\

\noindent
\begin{remark}\label{rem:controlenergy}
Notice that we denote by "control energy" the integral $\frac{1}{2}\int_{0}^{t_{f}}{\bf u}^{T}{\bf u}~dt$
~or~ $\frac{1}{2}\int_{0}^{t_{f}}{\dot{\bf v}}_{a}^{T} {\dot{\bf v}}_{a}dt$.
The question of whether this expression corresponds to the ``true fuel consumption'',
or "true energy'', is a discussion presented in different papers
such as \cite{Guo2019}, where the authors state that such quadratic
expression is meaningful because it is very difficult to handle analytically
a "true energy'' consumption formulation. They also argue that this cost function provides
a near-fuel-optimal steering law for a wide variety of problems, such as the lunar-landing guidance methods \cite{2019PingLu}.

\end{remark}


\section{Problem Solution}
\label{three}
\textcolor{black}{
\subsection{Rendez-Vous Case}}
\noindent
\begin{theorem}
The solution of the problem (\ref{eq:statement}) for the rendez-vous case is
\begin{equation}
{\bf u^{*}}=-{\bf p}_{v}(t)=t{\bf {\bf p}}_{r}-{\bf p}_{v}(0)\label{eq:acel1},
\end{equation}
where
\ifOneCol
    \begin{eqnarray}
    {\bf {\bf p}}_{r}=\dfrac{6\left[2({\bf r_{0}}-{\bf r_{f}})+({\bf v}_{g0}+{\bf v}_{gf}-\Delta_{{\bf wf}})t_{f}+2{\bf \varpi}_{f}\right]}{t_{f}^{3}}=
    \dfrac{6\left[2({\bf r_{0}}-{\bf r_{f}}+{\bf I_{wf}})+({\bf v}_{a0}+{\bf v}_{af})t_f\right]}{t_{f}^{3}},\nonumber\\
    {\bf {\bf p}}_{v}(0)=\dfrac{2\left[3({\bf r_{0}}-{\bf r_{f}})+(2{\bf v}_{g0}+{\bf v}_{gf}-\Delta_{{\bf wf}})t_{f}+3{\bf \varpi}_{f}\right]}{t_{f}^{2}}=
    \dfrac{2\left[3({\bf r_{0}}-{\bf r_{f}}+{\bf I_{wf}})+(2{\bf v}_{a0}+{\bf v}_{af})t_f\right]}{t_{f}^{2}},\nonumber\\
    \label{eq:JVD-1}
    \end{eqnarray}
\else
    \begin{eqnarray}
    {\bf {\bf p}}_{r}=\dfrac{6\left[2({\bf r_{0}}-{\bf r_{f}})+({\bf v}_{g0}+{\bf v}_{gf}-\Delta_{{\bf wf}})t_{f}+2{\bf \varpi}_{f}\right]}{t_{f}^{3}}\nonumber\\=
    \dfrac{6\left[2({\bf r_{0}}-{\bf r_{f}}+{\bf I_{wf}})+({\bf v}_{a0}+{\bf v}_{af})t_f\right]}{t_{f}^{3}},\nonumber\\
    {\bf {\bf p}}_{v}(0)=\dfrac{2\left[3({\bf r_{0}}-{\bf r_{f}})+(2{\bf v}_{g0}+{\bf v}_{gf}-\Delta_{{\bf wf}})t_{f}+3{\bf \varpi}_{f}\right]}{t_{f}^{2}}\nonumber\\=
    \dfrac{2\left[3({\bf r_{0}}-{\bf r_{f}}+{\bf I_{wf}})+(2{\bf v}_{a0}+{\bf v}_{af})t_f\right]}{t_{f}^{2}},\nonumber\\
    \label{eq:JVD-1}
    \end{eqnarray}
    \vspace{-15pt}
\fi

and

\ifOneCol
    \begin{equation}
    \begin{array}{c}
    {\bf I_w}(t)=\int_{0}^{t}{ {\bf w}}(\lambda)d\lambda,~{\bf I_{wf}}={\bf I_w}(t_f), ~~~~~~~~~~
    \Delta_{{\bf w}}(t)=\int_{0}^{t}{ {\dot{\bf w}}}(\lambda)d\lambda={\bf w}(t)-{\bf w}(0),\\
    {\bf \varpi}(t)=\int_{0}^{t}\Delta_{{\bf w}}(\tau)d\tau=\int_{0}^{t}\int_{0}^{\tau}{ {\dot{\bf w}}}(\lambda)d\lambda d\tau=\int_{0}^{t}{\bf w}(\lambda)d\lambda-{\bf w}(0)t.
    \end{array}\label{eq:deltaOmega}
    \end{equation}
\else
    \vspace{-5pt}
    \begin{equation}
    \begin{aligned}
    {\bf I_w}(t)&=\int_{0}^{t}{ {\bf w}}(\lambda)d\lambda,~{\bf I_{wf}}={\bf I_w}(t_f), \\
    \Delta_{{\bf w}}(t)&=\int_{0}^{t}{ {\dot{\bf w}}}(\lambda)d\lambda={\bf w}(t)-{\bf w}(0),\\
    {\bf \varpi}(t)&=\int_{0}^{t}\Delta_{{\bf w}}(\tau)d\tau=\int_{0}^{t}\int_{0}^{\tau}{ {\dot{\bf w}}}(\lambda)d\lambda d\tau\\
    &=\int_{0}^{t}{\bf w}(\lambda)d\lambda-{\bf w}(0)t.
    \end{aligned}\label{eq:deltaOmega}
    \end{equation}
\fi

The optimal cost is
\begin{equation}\label{optimalcost}
J^*=C_It_f+a_1t_f^{-1}+a_2t_f^{-2}+a_3t_f^{-3},
\end{equation}
where
\ifOneCol
    \begin{equation}
    a_1 = 2\left(\|{\bf v_{a0}}\|^{2}+\|{\bf v_{af}}\|^{2}+{\bf v_{a0}}^T{\bf v_{af}}\right),~~~~~
    a_2=6\left({\bf r_0}-{\bf r_f}+{\bf I_{wf}}\right)^T\left({\bf v_{a0}}+{\bf v_{af}}\right),~~~~~~
    a_3=6\|{\bf r_0}-{\bf r_f}+{\bf I_{wf}}\|^2.
    \end{equation}
\else
    \begin{equation}
    \begin{aligned}
    a_1 &= 2\left(\|{\bf v_{a0}}\|^{2}+\|{\bf v_{af}}\|^{2}+{\bf v_{a0}}^T{\bf v_{af}}\right),\\
    a_2&=6\left({\bf r_0}-{\bf r_f}+{\bf I_{wf}}\right)^T\left({\bf v_{a0}}+{\bf v_{af}}\right),\\
    a_3&=6\|{\bf r_0}-{\bf r_f}+{\bf I_{wf}}\|^2.
    \end{aligned}
    \end{equation}
\fi

The optimal flight times are the real positive roots of
\begin{equation}\label{polynomial}
P(t_f)=C_It_f^4-a_1t_f^2-2a_2t_f-3a_3
\end{equation}
for which the derivative of this polynomial with respect to $t_f$ is positive.
\end{theorem}
\noindent
\begin{proof}
The Hamiltonian function for the rendez-vous optimization problem (\ref{eq:statement}) is given by
\begin{equation}
H=C_{I}+\frac{1}{2}{\bf u}^{T}{\bf u}+{\bf p}_{r}^{T}{\bf v}_{g}+{\bf p}_{v}^{T}({ \dot{\bf w}+{\bf u}})\label{eq:Ham}
\end{equation}
where ${\bf p}_{r}$ and ${\bf p}_{v}$ are the costate vectors associated
with the position and velocity vectors, respectively.
A necessary condition for optimality is $H_{{\bf u}}=\frac{\partial H}{\partial{\bf u}}=0$,
which yields
$
{\bf u^{*}}= {\dot{\bf v}}_{a}^{*}=-{\bf p}_{v}\label{eq:ustar}
$.
According to Hamilton's equations the costate dynamics are
\begin{equation}
\begin{array}{ccc}
{ \dot{\bf p}}_{r}=-\dfrac{\partial H}{\partial{\bf r}}={\bf 0},
& \,\,\,\,\,\,\,\,\,\,\,\,\,&
{ \dot{\bf p}}_{v}=-\dfrac{\partial H}{\partial{\bf v_{g}}}={\bf -{\bf p}}_{r}
\end{array}
\end{equation}
Therefore we can write
\begin{equation}
{\bf p}_{v}(t)=-t{\bf {\bf p}}_{r}+{\bf p}_{v}(0)\label{eq:Jvtt}
\end{equation}
Thus, the optimal acceleration relative to the air can be written as in equation (\ref{eq:acel1}).
From (\ref{eq:acel1}) and (\ref{eq:model}), the resulting optimal
inertial acceleration is given by
\begin{equation}
{ {\dot{\bf{v}}^{*}}}_{g}(t)={\bf u^{*}}+{ {\dot{\bf w}}=}t{\bf {\bf p}}_{r}-{\bf p}_{v}(0)+{ {\dot{\bf w}}}
\end{equation}
Thus, integrating the optimal acceleration ${\dot{\bf{v}}^{*}}_{g}(t)$ twice yields
\begin{equation}
\begin{array}{c}
{{\bf v}^*}_{g}(t)={\bf v}_{g0}+\frac{t^{2}}{2}{\bf {\bf p}}_{r}-t{\bf p}_{v}(0)+\int_{0}^{t}{{\dot{\bf w}}}(\lambda)d\lambda\\
{\bf r^*}(t)={\bf r}_{0}+{\bf v}_{g0}t+\frac{t^{3}}{6}{\bf {\bf p}}_{r}-\frac{t^{2}}{2}{\bf p}_{v}(0)+\int_{0}^{t}\int_{0}^{\tau} {\dot{ \bf w}}(\lambda)d\lambda d\tau
\end{array}
\label{eq:X1s}
\end{equation}
or, equivalently,
\begin{equation}
\begin{array}{c}
{\bf v^*}_{g}(t)={\bf v}_{g0}+\frac{t^{2}}{2}{\bf {\bf p}}_{r}-t{\bf p}_{v}(0)+({\bf w}(t)-{\bf w}(0))\\
{\bf r^*}(t)={\bf r}_{0}+{\bf v}_{g0}t+\frac{t^{3}}{6}{\bf {\bf p}}_{r}-\frac{t^{2}}{2}{\bf p}_{v}(0)+\int_{0}^{t}{\bf w}(\lambda)d\lambda-{\bf w}(0)t
\end{array}\label{eq:Xs}
\end{equation}
Let us define the wind-related variables (\ref{eq:deltaOmega}).
Additionaly, for ease of notation, let us define $\Delta_{{\bf w}}(t_{f})=\Delta_{{\bf wf}}$
and ${\bf \varpi}(t_{f})={\bf \varpi}_{f}$.
Then, using (\ref{eq:Xs}) evaluated for the final time $t_{f}$,
we can solve a set of equations to find the optimal costate vectors as expressed in (\ref{eq:JVD-1}).
Therefore, the optimal control acceleration from (\ref{eq:acel1}) can be rewritten using (\ref{eq:JVD-1}) as
\ifOneCol
    \begin{equation}\label{optimalu}
    {\bf u}^{*}(t)=\frac{2}{t_f^3}\left[\left({\bf r}_0-{\bf r}_f+{\bf I_{wf}}\right)\left(6t-3t_f\right)+{\bf v}_{a0}t_f(3t-2t_f)+{\bf v}_{af}t_f(3t-t_f)\right]
    \end{equation}
\else
    \begin{multline}\label{optimalu}
    {\bf u}^{*}(t)=\frac{2}{t_f^3}[({\bf r}_0-{\bf r}_f+{\bf I_{wf}})(6t-3t_f)\\+{\bf v}_{a0}t_f(3t-2t_f)+{\bf v}_{af}t_f(3t-t_f)]
    \end{multline}
\fi
Using expression (\ref{optimalu}) the optimal cost can be obtained by computing the integral in (\ref{eq:statement}), which yields (\ref{optimalcost}).
Finally, to determine the optimal flight times one can simply take the derivative of (\ref{optimalcost}) relative to $t_f$ and equate it to zero.
The derivative of the optimal cost with respect to $t_f$ multiplied by $t_f^4$ yields the polynomial (\ref{polynomial}).
Therefore, since $t_f^4>0$ for $t_f\neq 0$, the real positive roots of (\ref{polynomial}) for which the derivative of the polynomial (\ref{polynomial}) with respect to $t_f$ is positive will be the optimal flight times, which finishes the proof.
\end{proof}

\noindent
\begin{remark}\label{rem:generalization_2}
This theorem generalizes the results from Yuan and Rodrigues \cite{YuanRodrigues2019}, who considered the situation of zero wind but with a constant gravitational acceleration. It is interesting to note that the results for the scenario of constant wind in this paper (including no wind) corresponds to the scenario of no gravitational acceleration in reference \cite{YuanRodrigues2019}, given that the role of $\dot{\bf w}$ is replaced by the gravitational acceleration in \cite{YuanRodrigues2019}.
Furthermore, when the wind is constantly accelerating so that $\dot {\bf w}={\bf k}$, then the results of this paper coincide with the ones in reference \cite{YuanRodrigues2019} for the particular case of ${\bf k}=[0~0~-g]^T$, where $g$ is the magnitude of the gravitational acceleration.
\end{remark}


\noindent
\begin{remark}\label{rem:wind}
Note that the term ${\bf r}_0-{\bf r}_f+{\bf I_{wf}}=-\int_0^{t_f}{\bf v_{a}}(\tau)d\tau$ appearing in the expressions for the coefficients $a_1, a_2, a_3$ of the cost (\ref{optimalcost}) does not depend explicitly on the final value of the wind ${\bf w}_f$. Therefore, the optimal cost does not explicitly depend on ${\bf w}_f$. The optimal control (\ref{optimalu}) does also not explicitly depend on ${\bf w}_f$, for the same reason. 
For constantly accelerating wind of the form $\dot{\bf w}={\bf k}$, or ${\bf w}(t)={\bf w}_0+{\bf k}t$, one has ${\bf I_{wf}}={\bf w}_0t_f+0.5{\bf k}t_f^2$ and
\ifOneCol
    \begin{equation}\label{secondu}
    {\bf u}^{*}(t)=\frac{2}{t_f^3}\left[\left({\bf r}_0-{\bf r}_f\right)\left(6t-3t_f\right)+{\bf v}_{g0}t_f(3t-2t_f)+{\bf v}_{gf}t_f(3t-t_f)\right]-{\bf k}
    \end{equation} 
\else
    \begin{multline}\label{secondu}
     {\bf u}^{*}(t)=\frac{2}{t_f^3}[({\bf r}_0-{\bf r}_f)(6t-3t_f)+{\bf v}_{g0}t_f(3t-2t_f)\\+{\bf v}_{gf}t_f(3t-t_f)]-{\bf k}
    \end{multline} 
\fi
The case of constant wind corresponds to making ${\bf k}=0$. 
\end{remark}

\vspace{-20pt}
\textcolor{black}{
\subsection{Intercept Case}}
Now, we will present the optimal solution for the Intercept case, where the objective is to find the optimal acceleration command to
make the UAV reach the target point at a free terminal velocity, instead of a constrained one. 
This problem is similar to the one presented in \cite{Bakolas2014}, which also considers the wind influence. However, instead of "minimum time" only, we use a generalized performance index and, further,
the control input considered here is not forced to be constrained.


Let the motion of the UAV be described by the same set of vector equations
presented previously in (\ref{eq:model}).
For the optimal intercept problem we have now a free ${\bf v}_g(t_f)$.
We consider here that the wind dynamics is given by ${ {\dot{ \bf w}}}(t)={\bf k}$.
The Hamiltonian function is then given by
\begin{equation}
H=C_{I}+\frac{1}{2}{\bf u}^{T}{\bf u}+{\bf p}_{r}^{T}{\bf v}_{g}+{\bf p}_{v}^{T}({\bf {{\bf k}+u}})\label{eq:Ham-1}
\end{equation}
where ${\bf p}_{r}$ and ${\bf p}_{v}$ are the costate vectors associated
with the position and velocity, respectively. Note that if
the terminal velocity is free, then the optimization problem
requires the terminal velocity costate to be zero, i.e., ${\bf p}_{v}(t_{f})={\bf 0}$.
Since ${\bf p}_{v}(t_{f})={\bf 0}$, from equation (\ref{eq:Jvtt}) and (\ref{eq:acel1}), we can write
\begin{equation}
{\bf p}_{v}(t)=(t_{f}-t){\bf {\bf p}}_{r},~~~~~~~~~~~
{\bf u^{*}=}~(t-t_{f}){\bf {\bf p}}_{r}
\label{eq:acel1-2}
\end{equation}
From (\ref{eq:model}) and (\ref{eq:acel1-2}) the optimal inertial acceleration is given by
\begin{equation}\label{dotvgintercept}
{ {\dot{\bf v}^{*}}}_{g}(t)={\bf u^{*}}+{\bf {\bf k}=}~(t-t_{f}){\bf {\bf p}}_{r}+{\bf {\bf k}}
\end{equation}
Successive integrations of this expression yield
\begin{equation}
\begin{array}{c}
{\bf v}_{g}(t)={\bf v}_{g0}+\frac{t^{2}}{2}{\bf {\bf p}}_{r}-t_{f}t{\bf p}_{r}+{\bf k}t\\
{\bf r}(t)={\bf r}_{0}+{\bf v}_{g0}t+\frac{t^{3}}{6}{\bf {\bf p}}_{r}-\frac{t^{2}}{2}t_{f}{\bf p}_{r}+{\bf k}\frac{t^{2}}{2}
\end{array}\label{eq:Xs-1}
\end{equation}
Evaluating these equations for $t_{f}$ the corresponding
optimal costates and optimal acceleration input can be written as
\ifOneCol
    \begin{equation}
    {\bf {\bf p}}_{r} = \dfrac{3\left[{\bf \Delta_r}+{\bf v}_{g0}t_{f}+{\bf k}\frac{t_{f}^{2}}{2}\right]}{t_{f}^{3}},\label{printercept}~~~~
    {\bf {\bf p}}_{v}(0) =
    \dfrac{3\left[{\bf \Delta_r}+{\bf v}_{g0}t_{f}\right]}{t_{f}^{2}}+\frac{3}{2}{\bf k},~~~~
    {\bf u}^{*}(t) =
    (t-t_{f})\dfrac{3\left[{\bf \Delta_r}+{\bf v}_{g0}t_{f}+{\bf k}\frac{t_{f}^{2}}{2}\right]}{t_{f}^{3}}.
    \end{equation}
\else
    \begin{align}
    {\bf {\bf p}}_{r} &= \dfrac{3\left[{\bf \Delta_r}+{\bf v}_{g0}t_{f}+{\bf k}\frac{t_{f}^{2}}{2}\right]}{t_{f}^{3}},\label{printercept} \nonumber \\
    {\bf {\bf p}}_{v}(0) &=
    \dfrac{3\left[{\bf \Delta_r}+{\bf v}_{g0}t_{f}\right]}{t_{f}^{2}}+\frac{3}{2}{\bf k},\nonumber \\
    {\bf u}^{*}(t) &=
    (t-t_{f})\dfrac{3\left[{\bf \Delta_r}+{\bf v}_{g0}t_{f}+{\bf k}\frac{t_{f}^{2}}{2}\right]}{t_{f}^{3}}.
    \end{align}
\fi
The initial and final acceleration input values are given by
\begin{equation}
\begin{array}{ccc}
{\bf u}^{*}(0)=-3\dfrac{\left[{\bf \Delta_r}+{\bf v}_{g0}t_{f}\right]}{t_{f}^{2}}-\frac{3}{2}{\bf k}, & \,\,\,\,\,\,\,\,\,\,\,\, & {\bf u}^{*}(t_{f})=0\end{array}
\end{equation}
From (\ref{eq:Xs-1}) and (\ref{printercept}) the final inertial velocity is written as
\begin{equation}
{\bf v}_{g}(t_{f})=
{\bf v}_{g0}-\frac{3}{2}\dfrac{\left[{\bf \Delta_r}+{\bf v}_{g0}t_{f}\right]}{t_{f}}+\frac{1}{4}{\bf k}t_{f}
\end{equation}
Computing now the optimal cost $J$ and $P(t_f)$ using the same methodology used in the rendez-vous case  yields
\ifOneCol
    \begin{equation}
    \begin{array}{c}
    J=\left(C_I+\frac{3}{8}{\bf k}^T{\bf k}\right)t_f+\frac{3}{2}{\bf v}_{g0}^T{\bf k}+\frac{3}{2}\left(\|{\bf v}_{g0}\|^2+{\bf\Delta_r}^T{\bf k}\right)t_f^{-1}
    +3{\bf \Delta_r}^T{\bf v}_{g0}t_f^{-2}+\frac{3}{2}\|{\bf \Delta_r}\|^2t_f^{-3}
    \\ \\
    P(t_{f})=\left(C_I+\frac{3}{8}{\bf k}^T{\bf k}\right)t_{f}^{4}-\frac{3}{2}\left(\|{\bf v}_{g0}\|^2+{\bf\Delta_r}^T{\bf k}\right)t_f^2
    -6{\bf \Delta_r}^{T}{\bf v}_{g0}t_{f}-\frac{9}{2}{\bf \Delta_r}^{T}{\bf \Delta_r}
    \end{array}
    \end{equation}
\else
    \begin{equation}
    \begin{array}{c}
    J=\left(C_I+\frac{3}{8}{\bf k}^T{\bf k}\right)t_f+\frac{3}{2}{\bf v}_{g0}^T{\bf k}+\frac{3}{2}\left(\|{\bf v}_{g0}\|^2+{\bf\Delta_r}^T{\bf k}\right)t_f^{-1}\\
    +3{\bf \Delta_r}^T{\bf v}_{g0}t_f^{-2}+\frac{3}{2}\|{\bf \Delta_r}\|^2t_f^{-3}
    \\ \\
    P(t_{f})=\left(C_I+\frac{3}{8}{\bf k}^T{\bf k}\right)t_{f}^{4}-\frac{3}{2}\left(\|{\bf v}_{g0}\|^2+{\bf\Delta_r}^T{\bf k}\right)t_f^2\\
    -6{\bf \Delta_r}^{T}{\bf v}_{g0}t_{f}-\frac{9}{2}{\bf \Delta_r}^{T}{\bf \Delta_r}
    \end{array}
    \end{equation}
\fi
whose roots are in agreement with the ZEM/ZEV result for intercept guidance
from \cite{Hawking2013}, although in that work the term ${\bf k}$  corresponds to the gravity acceleration, instead of the wind
acceleration.

It is interesting to note that, for a constant wind (${\bf k=0}$) and with ${\bf r}_f=0$
this polynomial will be further reduced to
\begin{equation}
P(t_{f})=C_{I}t_{f}^{4}-\frac{3}{2}{\bf v}_{g0}{}^{T}{\bf v}_{g0}t_{f}^{2}-6{\bf r_{0}}^{T}{\bf v}_{g0}t_{f}-\frac{9}{2}{\bf r_{0}}{}^{T}{\bf r_{0}}\label{eq:poly_Inter}
\end{equation}

\section{Particular cases for different wind profiles}
\label{four}

In this section we analyze the optimal solution of the rendez-vous guidance problem
(\ref{eq:statement}) for two particular cases of wind conditions. We also propose, in the third part of the section, a guideline for the solution of the general wind case model.

\subsection{\textcolor{black}{Constant wind speed} with ${\bf r}_f=0, {\bf v}_{gf}=0$}
Note that when the wind is constant (including the case of zero wind) we have $\Delta_{{\bf w}}(t_{f})={\bf 0},\,\,{\bf \varpi}(t_{f})={\bf 0}$
from (\ref{eq:deltaOmega}).
Assuming the target at the origin (${\bf r_{f}}=0$) and a zero final velocity (${\bf v}_{gf}=0$),  the  optimal costates (\ref{eq:JVD-1}), acceleration (\ref{eq:acel1}), cost (\ref{optimalcost}), and polynomial (\ref{polynomial}) become
\begin{equation}
\begin{array}{ccc}
{\bf {\bf p}}_{r}=\dfrac{6\left[2{\bf r_{0}}+{\bf v}_{g0}t_{f}\right]}{t_{f}^{3}}, & \,\,\,\,\,\,\,\,\,\,\,\,\,\, & {\bf {\bf p}}_{v}(0)=\dfrac{2\left[3{\bf r_{0}}+2{\bf v}_{g0}t_{f}\right]}{t_{f}^{2}}\end{array}\label{eq:JVD-1-1-1}
\end{equation}
\begin{equation}
{\bf u}^{*}(t)=t{\bf {\bf p}}_{r}-{\bf p}_{v}(0)=6t\dfrac{\left[2{\bf r_{0}}+{\bf v}_{g0}t_{f}\right]}{t_{f}^{3}}-\dfrac{2\left[3{\bf r_{0}}+2{\bf v}_{g0}t_{f}\right]}{t_{f}^{2}}\label{eq:optbruno}
\end{equation}

\begin{equation}
J=C_It_f+\bar a_1t_f^{-1}+6{\bf \Delta_r}^T\left({\bf v}_{g0}+{\bf v}_{gf}\right)t_f^{-2}+6\|{\bf \Delta_r}\|^2t_f^{-3}
\end{equation}
\begin{equation}\label{polynomialwzero}
P(t_{f})=C_{I}t_{f}^{4}-\bar a_1t_{f}^{2}-12{\bf \Delta_r}^{T}\left({\bf v}_{g0}+{\bf v}_{gf}\right)t_{f}-18{\bf \Delta_r}^{T}{\bf \Delta_r}
\end{equation}
where 
\begin{align}
\bar a_1=2\left[\|{\bf v}_{g0}\|^2+\|{\bf v}_{gf}\|^2+{\bf v}_{g0}^T{\bf v}_{gf}\right]=2\|{\bf v}_{g0}\|^2, \\
~{\bf \Delta_r}={\bf r}_0-{\bf r}_f={\bf r}_0,
\end{align}
since ${\bf r}_f={\bf v}_{gf}=0$.


\subsection{Constant wind acceleration with ${\bf r}_f=0$}\label{fourtwo}
We consider now the case of a constant rate of the wind speed $\dot{ \bf w}(t)={\bf k}$.
It is also assumed that the terminal velocity ${\bf v}_{g}(t_f)$ may be
constrained to any desired value. 
The values of $\Delta_{{\bf w}}(t_{f})$ and ${\bf \varpi}(t_{f})$
can be first computed from (\ref{eq:deltaOmega}) for this case yielding
\begin{equation}
\begin{array}{c}
\Delta_{{\bf w}}(t_{f})=\int_{0}^{t_f}{ {\dot{\bf w}}}(t)dt={\bf w}(t_{f})-{\bf w}(0)={\bf k}t_{f}\\
{\bf \varpi}(t_{f})=\int_{0}^{t_f}\int_{0}^{\tau}{ {\dot{\bf w}}}(\lambda)d\lambda d\tau=\int_{0}^{t_{f}}{\bf w}(t)dt-{\bf w}(0)t_{f}=\frac{{\bf k}t_{f}^{2}}{2}
\end{array}\label{eq:deltaOmega2-1-1}
\end{equation}
Assuming ${\bf r}_{f}=0$, the  optimal costates (\ref{eq:JVD-1}), acceleration (\ref{eq:acel1}), cost (\ref{optimalcost}), and polynomial (\ref{polynomial}) become
\ifOneCol
    \begin{align}
    {\bf {\bf p}}_{r}=6\dfrac{\left[2{\bf r_{0}}+({\bf v}_{g0}+{\bf v}_{gf})t_{f}-{\bf k}t_{f}^{2}+2\frac{{\bf k}t_{f}^{2}}{2}\right]}{t_{f}^{3}}=6\dfrac{\left[2{\bf r_{0}}+({\bf v}_{g0}+{\bf v}_{gf})t_{f}\right]}{t_{f}^{3}}\\
    {\bf {\bf p}}_{v}(0)=\dfrac{2\left[3{\bf r_{0}}+(2{\bf v}_{g0}+{\bf v}_{gf}-{\bf k}t_{f})t_{f}+3\frac{{\bf k}t_{f}^{2}}{2})\right]}{t_{f}^{2}}=\dfrac{2\left[3{\bf r_{0}+}(2{\bf v}_{g0}+{\bf v}_{gf})t_{f}\right]}{t_{f}^{2}}+{\bf k}
    \end{align}
    \begin{equation}
    {\bf u}^{*}(t)=t{\bf {\bf p}}_{r}-{\bf p}_{v}(0)=6t\dfrac{\left[2{\bf r_{0}}+({\bf v}_{g0}+{\bf v}_{gf})t_{f}\right]}{t_{f}^{3}}-\dfrac{2\left[3{\bf r_{0}+}(2{\bf v}_{g0}+{\bf v}_{gf})t_{f}\right]}{t_{f}^{2}}-{\bf k}\label{eq:acc_comp}
    \end{equation}
\else
    \begin{align}
    {\bf {\bf p}}_{r}&=6\dfrac{\left[2{\bf r_{0}}+({\bf v}_{g0}+{\bf v}_{gf})t_{f}-{\bf k}t_{f}^{2}+2\frac{{\bf k}t_{f}^{2}}{2}\right]}{t_{f}^{3}}\nonumber\\&=6\dfrac{\left[2{\bf r_{0}}+({\bf v}_{g0}+{\bf v}_{gf})t_{f}\right]}{t_{f}^{3}}\\
    {\bf {\bf p}}_{v}(0)&=\dfrac{2\left[3{\bf r_{0}}+(2{\bf v}_{g0}+{\bf v}_{gf}-{\bf k}t_{f})t_{f}+3\frac{{\bf k}t_{f}^{2}}{2})\right]}{t_{f}^{2}}\nonumber\\&=\dfrac{2\left[3{\bf r_{0}+}(2{\bf v}_{g0}+{\bf v}_{gf})t_{f}\right]}{t_{f}^{2}}+{\bf k}
    \end{align}
    \begin{align}
    &{\bf u}^{*}(t)=t{\bf {\bf p}}_{r}-{\bf p}_{v}(0)\nonumber \\
    &=6t\dfrac{\left[2{\bf r_{0}}+({\bf v}_{g0}+{\bf v}_{gf})t_{f}\right]}{t_{f}^{3}}-\dfrac{2\left[3{\bf r_{0}+}(2{\bf v}_{g0}+{\bf v}_{gf})t_{f}\right]}{t_{f}^{2}}-{\bf k}\label{eq:acc_comp}
    \end{align}
\fi

\begin{equation}
J={\bf k}^T\left({\bf v}_{g0}-{\bf v}_{gf}\right)+b_4t_f-b_2t_f^{-1}-\frac{b_1}{2}t_f^{-2}-\frac{b_0}{3}t_f^{-3}
\label{eq:polyrendezvous1}
\end{equation}
\begin{equation}
P(t_{f})=b_4t_f^4+b_2t_f^2+b_1t_f+b_0
\label{eq:polyrendezvous2}
\end{equation}
where
\ifOneCol
    \begin{equation}
    b_4=C_{I}+\frac{1}{2}{\bf k}^{T}{\bf k},~~~ 
    b_2=-2\left[\|{\bf v}_{g0}\|^2+\|{\bf v}_{gf}\|^2+{\bf v}_{gf}^T{\bf v}_{g0}\right], ~~~
    b_1=-12{\bf \Delta_r}^T\left[{\bf v}_{g0}+{\bf v}_{gf}\right],~~~
    b_0=-18{\bf \Delta_r}^T{\bf \Delta_r}
    \label{eq:polygeral}
    \end{equation}
\else
    \begin{align}
    b_4&=C_{I}+\frac{1}{2}{\bf k}^{T}{\bf k},\nonumber \\
    b_2&=-2\left[\|{\bf v}_{g0}\|^2+\|{\bf v}_{gf}\|^2+{\bf v}_{gf}^T{\bf                 v}_{g0}\right],\nonumber \\
    b_1&=-12{\bf \Delta_r}^T\left[{\bf v}_{g0}+{\bf v}_{gf}\right],\nonumber \\
    b_0&=-18{\bf \Delta_r}^T{\bf \Delta_r}
    \label{eq:polygeral}
    \end{align}
\fi
with ${\bf\Delta_r}={\bf r}_0$ since ${\bf r}_f=0$.
Thus, we conclude that a linear time-varying wind works as an effective increase in the trade-off parameter $C_{I}$, due to the term ${\bf k}^T{\bf k}$. 
\textcolor{black}{Notice that, for a zero or constant wind (${\bf k=0}$), this optimal polynomial is the same quartic polynomial appearing in \cite{DSouza1997},  \cite{Lewis1991}, \cite{2017Kumar}, \cite{Carvalho2018} and \cite{YuanRodrigues2019}, when the wind is not considered.
In this sense, our solution is an extension of the  classical Linear Quadratic Minimum-Time problem (LQMT)
to the case of relative velocities/accelerations \cite{2017Kumar}.
}

\subsection{\textcolor{black}{Guidelines for the general wind case model}}

 \textcolor{black}{
 From the equations (\ref{eq:polyrendezvous2}) and (\ref{eq:polygeral}), we can note that different wind acceleration vectors that have the same norm ($\Vert {\bf k} \Vert$) yield the same flight time, because the polynomial coefficients depend only on ${\bf k}^{T}{\bf k}$. The corresponding cost-to-go $J$, however, will depend on the elements of ${\bf k}$, as from (\ref{eq:polyrendezvous1}), we have}
 \textcolor{black}{
 \begin{equation}
 \begin{array}{l}
 \scalebox{0.9}{$
 J= {\bf k}^T\left({\bf v}_{g0}-{\bf v}_{gf}\right)+(C_{I}+\frac{1}{2}{\bf k}^{T}{\bf k})t_f-b_2t_f^{-1}-\frac{b_1}{2}t_f^{-2}-\frac{b_0}{3}t_f^{-3}
 $}
 \end{array}
 \label{eq:J_corrected}
 \end{equation}
 }

 \textcolor{black}{
If we analyse the cost-to-go values for a given set of parameters $C_I,{\bf v}_{g0},{\bf v}_{gf},t_f$, we conclude that the sign of  ${\bf k}^{T}({\bf v}_{g0}-{\bf v}_{gf})$ is important.
For example,  ${\bf k}^{T}({\bf v}_{g0}-{\bf v}_{gf})<0$, in the above equation, will result in a lower cost-to-go $J$. This result suggests a methodology to guide the aircraft along a trajectory in a variable wind profile, under the assumption that the wind model can be approximated by a series of piecewise linear time-varying speed regions  (Fig. 2). The idea is to divide the longitudinal segment (x-axis) between the vehicle and the target into a finite number of small segments where, in each segment, the wind speed can be approximated by a linear time varying function, or ${{\bf w}}_i(t)={\bf w}_{i0}+{\bf k}_it$, where the control input from equation (\ref{secondu}) can then be applied.
}



   \begin{figure}[hbt]
\noindent \centering{}
\ifOneCol
    \includegraphics[scale=0.54]{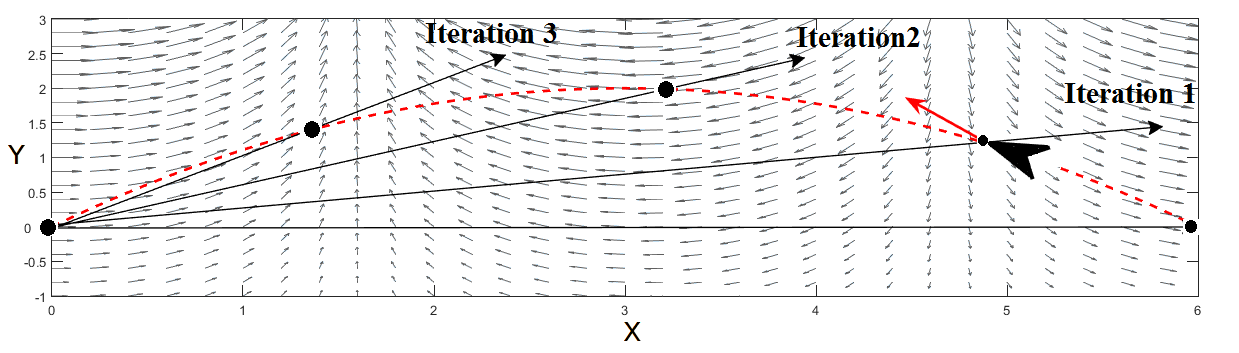}\caption{Iterative solution proposal for the general wind case.  }
\else
    \includegraphics[width=\columnwidth]{figs/traj_geral2.png}\caption{Iterative solution proposal for the general wind case.  }
\fi
\label{Figurejardin}
\end{figure}

 \textcolor{black}{
Further, inside each segment, we use a modified trade-off parameter as
 \begin{equation}
 \begin{array}{l}
 \scalebox{1}{$
C'_{I}=C_{I}+{\bf k}^T\left({\bf v}_{g0}-{\bf v}_{gf}\right)
 $}
 \end{array}
 \label{eq:J_corrected2}
 \end{equation}
 }
 
 \textcolor{black}{
Thus, when ${\bf k}^T\left({\bf v}_{g0}-{\bf v}_{gf}\right)>0$, the effective trade-off parameter will be increased to favor the  vehicle acceleration, and when it is negative, the trade-off will be decreased to decelerate the vehicle.}
    \textcolor{black}{
 Note also that, in the "minimum-time" case, or when $C_I$ is too large, the wind acceleration will make no difference, as $C_{I} \gg{\bf k}^T\left({\bf v}_{g0}-{\bf v}_{gf}\right)$.}

\section{Optimal solutions using ZEM/ZEV formulation}
\label{five}
\textcolor{black}{
\subsection{Optimal rendez-vous using ZEM/ZEV}}
The optimal solution for the general rendez-vous problem can also be obtained in an equivalent form using the Zero-Effort-Miss/Zero-Effort-Velocity (ZEM/ZEV) feedback
guidance approach in the presence of gravitational terms \cite{Hawking2013}. This technique is commonly used in aerospace applications and it has proven benefits for autonomous onboard implementation due to its feedback nature and the flexibility to consider a nonlinear model for gravitational or even atmospheric drag terms. However, when such terms are included, numerical integration methods are typically used to find a suboptimal solution  \cite{Furfaro2017}.
In this formulation, proposed in \cite{Ebrahimi2008}, the  zero-effort-velocity (ZEV)
and the zero-effort-miss (ZEM), which are used in classical guidance problems, are defined as
\begin{equation}
{\bf {\bf ZEM=}r}_{f}-\bar{{\bf r}}_{f},\,\,\,\,\,\,\,\,\,\,\,\,
 {\bf ZEV}={\bf v}_{gf}-{\bf \bar{v}}_{gf}
\end{equation}
where $\bar{{\bf r}}_{f}$ is the final interceptor (UAV) position without corrective action, and $\bar{{\bf v}}_{gf}$ is the final interceptor velocity without corrective action.
In summary, ${\bf ZEM}$ is the final position error or the miss error, and ${\bf ZEV}$ is the corresponding miss velocity. The ${\bf ZEM}/{\bf ZEV}$ errors are usually calculated
as functions of the so-called time-to-go, which is the time it will take to
reach the terminal (target) state from the current state and is
defined as $t_{go}=t_{f}-t$. 
In the ZEM/ZEV formulation, the optimal acceleration is expressed
as a function of the terminal time as
\begin{equation}
{\bf u^{*}=}-{\bf p}_{v}(t_{go})=-t_{go}{\bf {\bf p}}_{r}-{\bf p}_{v}(t_{f})
\label{eq:uzz}
\end{equation}

We also need to compute the wind variables $\Delta_{{\bf wg}}(t_{go})$
and ${\bf \varpi_g}(t_{go})$ similar to (\ref{eq:deltaOmega}) but with the final value $w(t_f)$ used as reference instead of $w(0)$.
For a constant wind speed rate $\dot{\bf  w}(t)={\bf k}$ we get, with $\tau=t_f-t$, the integrals
\ifOneCol
    \begin{equation}
    \begin{array}{c}
    \Delta_{{\bf wg}}(t_{go})=\int_{0}^{t_{go}}\frac{d{\bf w}}{d\tau}d\tau=-{\bf k}t_{go},~~~~~~~~~~~
    {\bf \varpi_g}(t_{go})=\int_{0}^{t_{go}}{\bf \Delta_w}(\tau)d\tau=-{\bf k}\frac{t_{go}^2}{2}
    \end{array}
    \label{eq:deltaOmega2-1-1-1-1}
    \end{equation}
\else
    \begin{align}
    \begin{array}{c}
    \Delta_{{\bf wg}}(t_{go})=\int_{0}^{t_{go}}\frac{d{\bf w}}{d\tau}d\tau=-{\bf k}t_{go}, \\
    {\bf \varpi_g}(t_{go})=\int_{0}^{t_{go}}{\bf \Delta_w}(\tau)d\tau=-{\bf k}\frac{t_{go}^2}{2}
    \end{array}
    \label{eq:deltaOmega2-1-1-1-1}
    \end{align}
\fi



Thus, using (\ref{eq:uzz}), and integrating the optimal inertial acceleration $\dot{ \bf v}^{*}_{g}={\bf u}^{*}+{\dot{\bf  w}}$,
we can derive the general expressions for the position and velocity vectors, that are now functions of the time-to-go, as
\begin{equation}
\begin{array}{c}
{\bf v}_{g}(t_{go})={\bf v}_{gf}+\frac{t_{go}^{2}}{2}{\bf {\bf p}}_{r}+t_{go}{\bf p}_{v}(t_{f})+\Delta_{{\bf wg}}(t_{go})\\
{\bf r}(t_{go})={\bf r_{f}}-{\bf v}_{gf}t_{go}-\frac{t_{go}^{3}}{6}{\bf {\bf p}}_{r}-\frac{t_{go}^{2}}{2}{\bf p}_{v}(t_{f})-{\bf \varpi_g}(t_{go})
\end{array}\label{eq:Vas}
\end{equation}
From the endpoint condition at $t_{go}$ we can proceed as before to solve a system of linear equations to write
\ifOneCol
    \begin{equation}
    {\bf {\bf p}}_{r}=\dfrac{6\left[2\left({\bf r}(t_{go})-{\bf r_f}\right)+({\bf v}_{g}(t_{go})+{\bf v}_{gf})t_{go}\right]}{t_{go}^{3}},~~~~~~~~
    {\bf {\bf p}}_{v}(t_{f})=\dfrac{-6\left({\bf r}(t_{go})-{\bf r_f}\right)-2({\bf v}_{g}(t_{go})+2{\bf v}_{gf})t_{go}}{t_{go}^{2}}+{\bf k}.
    \label{eq:JVD-3}
    \end{equation}
\else
    \begin{align}
    {\bf {\bf p}}_{r}&=\dfrac{6\left[2\left({\bf r}(t_{go})-{\bf r_f}\right)+({\bf v}_{g}(t_{go})+{\bf v}_{gf})t_{go}\right]}{t_{go}^{3}}, \nonumber \\
    {\bf {\bf p}}_{v}(t_{f})&=\dfrac{-6\left({\bf r}(t_{go})-{\bf r_f}\right)-2({\bf v}_{g}(t_{go})+2{\bf v}_{gf})t_{go}}{t_{go}^{2}}+{\bf k}.
    \label{eq:JVD-3}
    \end{align}
\fi
Using $t_{go}=t_f-t$, we define the ${\bf ZEM}$ and ${\bf ZEV}$ errors as
\begin{equation}
\begin{array}{c}
{\bf {\bf ZEV}}={\bf v}_{gf}-{\bf \bar{v}}_{gf}=
{\bf v}_{gf}-\left[{\bf v}_{g}(t_{go})+{\bf k}t_{go}\right]
\\
{\bf ZEM}={\bf r}_f-\bar{\bf r}_f =
{\bf r}_f-{\bf r}(t_{go})-{\bf v}_{g}(t_{go})t_{go}-{\bf k}\frac{t_{go}^{2}}{2}
\end{array}
\end{equation}
and then rewrite the costates at (\ref{eq:JVD-3}) as
\ifOneCol
    \begin{equation}
    {\bf {\bf p}}_{r}= \frac{12}{t_{go}^3}\left[\frac{t_{go}}{2}{\bf ZEV}-{\bf ZEM}\right],~~~~~~~~~~~~
    {\bf {\bf p}}_{v}(t_{f}) = \frac{-6}{t_{go}^2}\left[\frac{2t_{go}}{3}{\bf ZEV}-{\bf ZEM}\right].\label{pvZZ}
    \end{equation}
\else
    \begin{align}
    {\bf {\bf p}}_{r}&= \frac{12}{t_{go}^3}\left[\frac{t_{go}}{2}{\bf ZEV}-{\bf ZEM}\right],\nonumber \\
    {\bf {\bf p}}_{v}(t_{f}) &= \frac{-6}{t_{go}^2}\left[\frac{2t_{go}}{3}{\bf ZEV}-{\bf ZEM}\right].\label{pvZZ}
    \end{align}
\fi
The resulting optimal airspeed acceleration, from (\ref{eq:uzz}) and (\ref{pvZZ}), then becomes
\begin{equation}\label{ZEMZEVrendezvous}
{\bf u}^{*}(t_{go})=\frac{6}{t_{go}^{2}}{\bf ZEM}-\frac{2}{t_{go}}{\bf ZEV}
\end{equation}
Notice that when $t_{go}=t_{f}$ (or $t=0$), then ${\bf u}^{*}(t_{go})$
yields the same expression  for ${\bf u}^{*}(0)$ derived in (\ref{eq:acc_comp}).

\textcolor{black}{
\subsection{Optimal intercept using ZEM/ZEV}}
\textcolor{black}{
For the solution of the optimal intercept
problem, using the ZEM/ZEV approach, we recall that the terminal velocity costate is zero as the final velocity is free, and from  (\ref{eq:acel1-2}) and (\ref{dotvgintercept}) we can write the following relations:}
\ifOneCol
    \begin{equation}
    \begin{array}{l}
    {\bf p}_{v}(t_{go})=(t_{f}-t){\bf {\bf p}}_{r}=t_{go}{\bf {\bf p}}_{r},~~~~~~~~~~~~
    {\bf u}^*(t_{go})=-{\bf p}_v(t_{go})=-t_{go}{\bf {\bf p}}_{r}\\
    {\bf v}_{g}(t_{go})=\frac{t_{go}^{2}}{2}{\bf {\bf p}}_{r}-t_{go}{\bf {\bf k}}+{\bf v}_{gf},~~~~~~~~~~~~
    {\bf r}(t_{go})=-\frac{t_{go}^{3}}{6}{\bf {\bf p}}_{r}+{\bf k}\frac{t_{go}^{2}}{2}-{\bf v}_{gf}t_{go}+{\bf r}_{f}
    \end{array}
    \end{equation}
\else
    \begin{align}
    {\bf p}_{v}(t_{go})&=(t_{f}-t){\bf {\bf p}}_{r}=t_{go}{\bf {\bf p}}_{r},\nonumber \\
    {\bf u}^*(t_{go})&=-{\bf p}_v(t_{go})=-t_{go}{\bf {\bf p}}_{r}, \nonumber \\
    {\bf v}_{g}(t_{go})&=\frac{t_{go}^{2}}{2}{\bf {\bf p}}_{r}-t_{go}{\bf {\bf k}}+{\bf v}_{gf},\nonumber \\
    {\bf r}(t_{go})&=-\frac{t_{go}^{3}}{6}{\bf {\bf p}}_{r}+{\bf k}\frac{t_{go}^{2}}{2}-{\bf v}_{gf}t_{go}+{\bf r}_{f}
    \end{align}
\fi
which together yield
\begin{equation}
\begin{array}{l}
{\bf {\bf p}}_{r}=\dfrac{3\left[{\bf r}(t_{go})-{\bf r}_f+{\bf v}(t_{go})t_{go}+{\bf k}\frac{t_{go}^{2}}{2}\right]}{t_{go}^{3}}\\ 
\end{array}
\end{equation}
Finally, for ${\bf r}_f=0$ the optimal acceleration command is obtained as
\ifOneCol
    \begin{equation}
    {\bf u}^{*}(t)=-t_{go}\dfrac{3\left[{\bf r}(t_{go})-{\bf r}_f+{\bf v}(t_{go})t_{go}+{\bf k}\frac{t_{go}^{2}}{2}\right]}{t_{go}^{3}}=-\frac{3}{t_{go}^{2}}{\bf r}(t_{go})-\frac{3}{t_{go}}{\bf v}(t_{go})-\frac{3}{2}{\bf k}
    \end{equation}
\else 
    \begin{multline}
            {\bf u}^{*}(t)=-t_{go}\dfrac{3\left[{\bf r}(t_{go})-{\bf r}_f+{\bf v}(t_{go})t_{go}+{\bf k}\frac{t_{go}^{2}}{2}\right]}{t_{go}^{3}}
            \\=-\frac{3}{t_{go}^{2}}{\bf r}(t_{go})-\frac{3}{t_{go}}{\bf v}(t_{go})-\frac{3}{2}{\bf k}
    \end{multline}
\fi
And with the definition of ${\bf ZEM}$ expressed as
\begin{equation}
{\bf ZEM}={\bf r}_{f}-\bar{{\bf r}}_{f}=-\left({\bf r}(t_{go})+{\bf v}(t_{go})t_{go}+\frac{{\bf k}t_{go}^{2}}{2}\right)
\end{equation}
Then the optimal acceleration can be written as
\begin{equation}
{\bf u}^{*}(t_{f})=\frac{3}{t_{go}^{2}}{\bf ZEM}
\end{equation}
\section{On the existence of solutions for the minimum control energy case}
\label{seven}
This section focuses on the analysis of the number of solutions of the minimum control energy case for $C_I=0$.
Firstly, it is important to present the physical
interpretations of the roots of the polynomial (\ref{polynomial}) as a function of the initial state.
For that, we will consider only the $x$-component, since the movement in each coordinate $x-y$ is independent, as from (\ref{eq:X1s}).
We also assume that $x_{0}v_{gx0}<0$, such that the initial ground velocity component in the x-coordinate is pointing toward the target. Then, we can define a reference time-constant as $t_r=-{x_{0}}/{v_{gx0}}$, such that the final time can always be written as a multiple of the reference time, or $t_{f}=Kt_{r}$, for an appropriate constant $K>0$.
The physical interpretation of the velocity and acceleration responses for the rendez-vous case is summarized in Table 1 for $K=1,2,3,$ and $K>3$. It is possible to show that, whenever $K>3$, a reversal
movement in the x-axis will occur  with a change of sign in the velocity curve during the movement \cite{Guo2012}.
\begin{center}
\begin{table*}[h]
\centering
\caption{Special cases of rendez-vous optimal responses for $K=1,2,3$ and $>3$.}
\begin{small}
\label{table:draglift1}
\begin{tabular}{cccccc}
\hline 
\hline 
$K$ & $C_I$  & terminal velocity & position $x(t)$ & velocity $v_{gx}(t)$ & acceleration $\dot{v}_{gx}(t)$
\tabularnewline
\hline 
~~~1 & $\forall$  & $v_{gxf}\neq0$ & linear  & constant & zero
\tabularnewline

~~~2 & $\forall$  & $v_{gxf}=0$ & quadratic & linear & constant \tabularnewline

~~~3 & $\forall$ & $v_{gxf}=0$  & cubic & quadratic   & linear, $\dot{v}_{gx}(t_{f})=0$ \tabularnewline
$>$ 3  & $\forall$ & $v_{gxf}=0$  & cubic & quadratic  & linear, $\dot{v}_{gx}(t_{f})\neq0$
\tabularnewline
\hline 
\hline
\end{tabular}
\end{small}
\end{table*}
\end{center}
\vspace*{-0.8cm}

\begin{figure}[t]
\noindent \begin{centering}
\ifOneCol
    \includegraphics[scale=0.64]{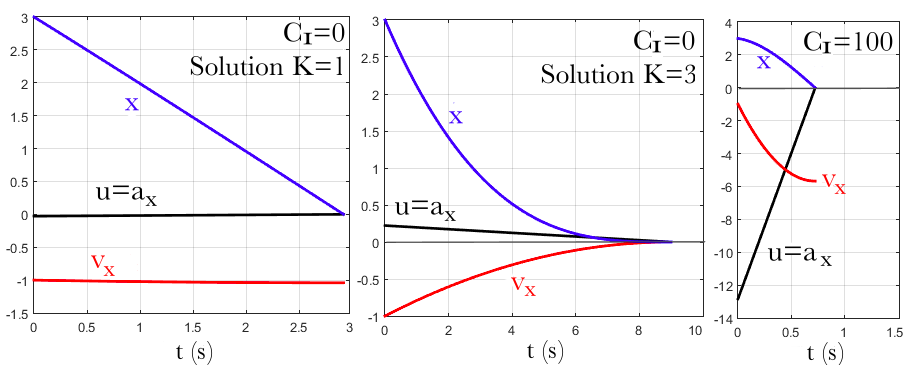}
\else
    \includegraphics[width=\columnwidth]{figs/intercept1d_modBruno2.png}
\fi
\par\end{centering}
\noindent \centering{}\caption{{Typical optimal intercept time responses for a UAV in constant wind, $x_{0}=3,\,v_{x0}=-1$
and  different $C_{I}.$}}
\label{interceptfigure}
\end{figure}

Regarding the intercept problem, typical plots of the optimal longitudinal acceleration and velocity ($x$-components), for constant wind, are illustrated in Figure \ref{interceptfigure}, for $C_I=0$ and $C_I=100$.
The two first plots, with $C_{I}=0$ correspond to the minimum control energy case. The last plot with $C_{I}=100$ is closer to the problem of minimum time ($C_I\to\infty$). 
Note that, in agreement with  \cite{Guo2012Generalz}, for the intercept with $C_{I}=0$ there are two feasible solutions: one with constant velocity $(K=1)$, and another with zero terminal acceleration $(K=3)$.

We will now analyze the number of optimal solutions for the minimum control energy case  $(C_I=0)$ in a 2D scenario. It has already been proved that the depressed quartic polynomial in (\ref{polynomial}) has always at least one positive real root when $C_I>0$ \cite{DSouza1997}, \cite{Carvalho2018}.
However, in the particular case of $C_I=0$, the quartic polynomial becomes a second order polynomial that may not admit a positive real root in some cases.
The existence of a feasible solution for this minimum energy problem $(C_I=0)$ is discussed in \cite{Guo2019} where a trade-off cost function is introduced in the ZEM/ZEV approach for orbital intercept/rendez-vous.
In that work, the Hamiltonian equation is very similar to (\ref{eq:Ham}), with the gravity vector in place of the wind acceleration vector.
According to the authors in \cite{Guo2019}, one possible explanation for the inexistence of solutions for $t_f$ is that, with $C_I=0$, the terminal time is neither constrained nor penalized.
However, this does not explain why there are some cases in which  $C_I=0$ still yields a finite positive terminal time, as we show in this section. 

\begin{figure}[t]
\noindent \begin{centering}
\ifOneCol
    \includegraphics[scale=0.58]{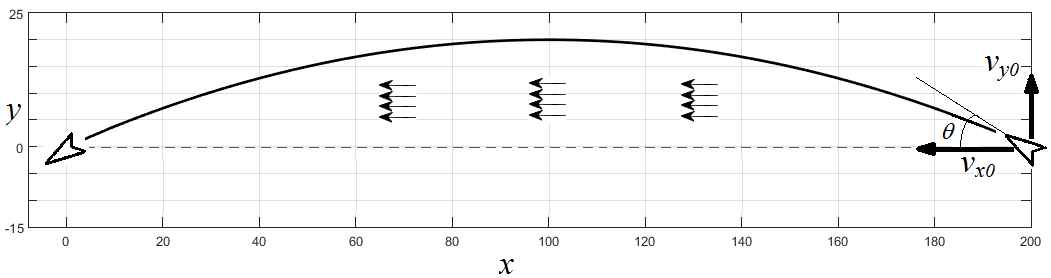}
\else
    \includegraphics[width=\columnwidth]{figs/AIAA_fig_wind_heading.png}
\fi
\par\end{centering}
\noindent \centering{}\caption{{Search for the maximum initial velocity heading $\theta$ with $C_I=0$ that still yields a feasible optimal $t_f$.}}
\label{nines}
\end{figure}

\textcolor{black}{
For a better understanding of this issue, we investigate two special cases, under a constant wind.
Consider the problem of a 2D path with $x(0)=x_0,y(0)=0, x_{0}v_{gx0}<0, C_I=0$, and ${\bf k=0}$, for both the rendez-vous and the intercept cases (Fig. \ref{nines}).
Suppose one wants to find the maximum initial velocity heading $\theta$ for which a finite terminal time $t_f$ exists. For the rendez-vous case, we assume that $v_{gxf}=v_{gx0}, v_{gyf}=-v_{gy0}$, and for the intercept case, the terminal velocity is free. }
Note that this problem is equivalent to one that searches for the maximum  allowable orthogonal velocity $v_{gy0}$, for a given fixed $v_{gx0}$, as the heading angle $\theta$ is given by $\theta=\arctan(\frac{v_{gy0}}{v_{gx0}})$.
The corresponding time optimal polynomials, using ${\bf r}_f=0$, in
expressions (\ref{eq:polyrendezvous1}) and (\ref{eq:poly_Inter}) are, respectively:

\ifOneCol
    \begin{equation}
    \begin{array}{l}
     \mbox{Rendez-vous}~~~~~   \\
     \color{black}{
    P(t_{f})=(C_{I}+\frac{1}{2}{\bf k}^{T}{\bf k})t_{f}^{4}
    -2({\bf v}_{g0}{}^{T}{\bf v}_{g0}+{\bf v}_{gf}{}^{T}{\bf v}_{gf}+{\bf v}_{gf}^{T}{\bf v}_{g0})t_{f}^{2}
    -12{\bf r_{0}}^{T}({\bf v}_{g0}+{\bf v}_{gf})t_{f}-18{\bf r_{0}}{}^{T}{\bf r_{0}}}\\
    P(t_{f})=C_{I}t_{f}^{4}
    -2(3v_{gx0}^{2}+v_{gy0}^{2})t_{f}^{2}
    -24(x_{0}v_{gx0})t_{f}-18x_{0}^{2}\\ 
    \mbox{For $C_I=0$}, P(t_{f})=0\to-(3v_{gx0}^{2}+v_{gy0}^{2})t_{f}^{2}-12(x_{0}v_{gx0})t_{f}-9x_{0}^{2}=0 \\ 
      \label{eq:poly_compara1b}
    \end{array}
    \end{equation}

    \begin{equation}
    \begin{array}{l}
    \mbox{Intercept}~~~~~~~~~~\\
       \color{black}{
      P(t_{f})=C_{I}t_{f}^{4}-\frac{3}{2}{\bf v}_{g0}{}^{T}{\bf v}_{g0}t_{f}^{2}-6{\bf r_{0}}^{T}{\bf v}_{g0}t_{f}-\frac{9}{2}{\bf r_{0}}{}^{T}{\bf r_{0}}}\\
      
      P(t_{f})=C_{I}t_{f}^{4}-\frac{3}{2}(v_{gx0}^{2}+v_{gy0}^{2})t_{f}^{2}-6(x_{0}v_{gx0})t_{f}-\frac{9}{2}x_{0}^{2}=0 \\
      
     \mbox{For $C_I=0$},   P(t_{f})=0\to-3(v_{gx0}^{2}+v_{gy0}^{2})t_{f}^{2}-12(x_{0}v_{gx0})t_{f}-9x_{0}^{2}=0 \\
      
      \label{eq:poly_compara1}
    \end{array}
    \end{equation}
\else
    \begin{equation}
    \begin{array}{l}
     \mbox{Rendez-vous}~~~~~   \\
     
    \color{black}{
    P(t_{f})=(C_{I}+\frac{1}{2}{\bf k}^{T}{\bf k})t_{f}^{4}}
    \\ \color{black}{
    ~~~~~~~~~~~~~~~~~-2({\bf v}_{g0}{}^{T}{\bf v}_{g0}+{\bf v}_{gf}{}^{T}{\bf v}_{gf}+{\bf v}_{gf}^{T}{\bf v}_{g0})t_{f}^{2}}
    \\  \color{black}{
    ~~~~~~~~~~~~~~~~~~~~~~~~~~~~-12{\bf r_{0}}^{T}({\bf v}_{g0}+{\bf v}_{gf})t_{f}-18{\bf r_{0}}{}^{T}{\bf r_{0}}}\\
    P(t_{f})=C_{I}t_{f}^{4}
    -2(3v_{gx0}^{2}+v_{gy0}^{2})t_{f}^{2}
    -24(x_{0}v_{gx0})t_{f}-18x_{0}^{2}\\ 
    \mbox{For $C_I=0$}, P(t_{f})=0 \\ ~~~~~~~~~~~~\to-(3v_{gx0}^{2}+v_{gy0}^{2})t_{f}^{2}-12(x_{0}v_{gx0})t_{f}-9x_{0}^{2}=0 \\ 
      \label{eq:poly_compara1b}
    \end{array}
    \end{equation}
     \vspace{-1cm}
     
    \vspace{-15pt}
    \begin{equation}
    \begin{array}{l}
    \mbox{Intercept}~~~~~~~~~~\\
       \color{black}{
      P(t_{f})=C_{I}t_{f}^{4}-\frac{3}{2}{\bf v}_{g0}{}^{T}{\bf v}_{g0}t_{f}^{2}-6{\bf r_{0}}^{T}{\bf v}_{g0}t_{f}-\frac{9}{2}{\bf r_{0}}{}^{T}{\bf r_{0}}}\\
      
      P(t_{f})=C_{I}t_{f}^{4}-\frac{3}{2}(v_{gx0}^{2}+v_{gy0}^{2})t_{f}^{2}
      \\~~~~~~~~~~~~~~~~~~~~~~~~~~~~~~~~~~~-6(x_{0}v_{gx0})t_{f}-\frac{9}{2}x_{0}^{2}=0 \\
      
     \mbox{For $C_I=0$},   P(t_{f})=0\\
     ~~~~~~~~~~~~\to-3(v_{gx0}^{2}+v_{gy0}^{2})t_{f}^{2}-12(x_{0}v_{gx0})t_{f}-9x_{0}^{2}=0 \\
      
      \label{eq:poly_compara1}
    \end{array}
    \end{equation}
\fi

Note that the resulting quadratic polynomials are very similar for both problems. If we now write the optimal terminal time $t_f$ as a function of the reference time (Tab. 1), that is, $t_{f}=K\left( -\frac{x_0}{v_{gx0}}\right)$, we can rewrite $P(t_f)$ as $P(K)$, or

\ifOneCol
    \begin{equation}
    \begin{array}{ll}
     \mbox{Rendez-vous} & \\
     &  P(K)=0\to-(3v_{gx0}^{2}+v_{gy0}^{2})K^{2}\left( -\frac{x_{g0}}{v_{gx0}} \right)^{2}-12(x_{0}v_{gx0})K\left( -\frac{x_{g0}}{v_{gx0}} \right)-9x_{0}^{2}=0 \\
    &  P(K)=0\to
      \left(\dfrac{-3(v_{gx0}^{2}+\frac{1}{3}v_{gy0}^{2})}{v_{gx0}^{2}}\right)    K^{2}x_{0}^{2}+12Kx_{0}^{2}-9x_{0}^{2}=\alpha_{0R}
      K^{2}+12K-9=0\\ 
        \label{eq:poly_compara2b}
    \end{array}
    \end{equation}
    \begin{equation}
    \begin{array}{ll}
      \mbox{Intercept} & \\
       & P(K)=0\to-3(v_{gx0}^{2}+v_{gy0}^{2})K^{2}\left( -\frac{x_{g0}}{v_{gx0}} \right)^{2}-12(x_{0}v_{gx0})K\left( -\frac{x_{g0}}{v_{gx0}} \right)-9x_{0}^{2}=0 \\
      &     P(K)=0\to
       \left(\dfrac{-3(v_{gx0}^{2}+v_{gy0}^{2})}{v_{gx0}^{2}}\right)
      K^{2}x_{0}^{2}+12Kx_{0}^{2}-9x_{0}^{2}=
      \alpha_{0I}   K^{2}+12K-9=0 
      \label{eq:poly_compara}
    \end{array}
    \end{equation}
\else
    \begin{equation}
    \begin{array}{ll}
     \mbox{Rendez-vous} \\
       P(K)=0\to-(3v_{gx0}^{2}+v_{gy0}^{2})K^{2}\left( -\frac{x_{g0}}{v_{gx0}} \right)^{2}
     \\~~~~~~~~~~~~~~~~~~~~~~~~~-12(x_{0}v_{gx0})K\left( -\frac{x_{g0}}{v_{gx0}} \right)-9x_{0}^{2}=0 \\
      P(K)=0\to
      \left(\dfrac{-3(v_{gx0}^{2}+\frac{1}{3}v_{gy0}^{2})}{v_{gx0}^{2}}\right)    K^{2}x_{0}^{2}
      \\~~~~~~~~~~~~+12Kx_{0}^{2}-9x_{0}^{2}=\alpha_{0R}
      K^{2}+12K-9=0
      \label{eq:poly_compara2c}
       \end{array}
    \end{equation}
        \begin{equation}
    \begin{array}{ll}
          \mbox{Intercept}  \\
        P(K)=0\to-3(v_{gx0}^{2}+v_{gy0}^{2})K^{2}\left( -\frac{x_{g0}}{v_{gx0}} \right)^{2}
       \\~~~~~~~~~~~~~~~~~~~~~~~~~-12(x_{0}v_{gx0})K\left( -\frac{x_{g0}}{v_{gx0}} \right)-9x_{0}^{2}=0 \\
          P(K)=0\to
       \left(\dfrac{-3(v_{gx0}^{2}+v_{gy0}^{2})}{v_{gx0}^{2}}\right)
      K^{2}x_{0}^{2}
      \\~~~~~~~~~~~~+12Kx_{0}^{2}-9x_{0}^{2}=
      \alpha_{0I}   K^{2}+12K-9=0 \\
     
      \label{eq:poly_compara}
    \end{array}
    \end{equation}
\fi

The roots of both polynomials can be real or complex. 
It is straightforward to see that if the initial velocity heading $\theta$ is zero, then there exists a positive real $t_f$ solution for the quadratic polynomial. Indeed, in this case, as $v_{gy0}=0$, both polynomials are equal to $P(K)=-3 K^{2}+12K-9=0$, whose roots are $K=1$ and $K=3$. 
Thus, we know that there exists a feasible solution for $P(K)=\alpha_0 K^{2}+12K-9=0$ when $v_{gy0}=0$.
As we increase $|v_{gy0}|$, for a given $v_{gx0}$, the corresponding polynomial coefficient $\alpha_0$ will change, as will the pair of roots, until they become a double real root, in the transition for a complex conjugate pair. In this critical limit, the polynomial discriminant  $(\Delta=144+36\alpha_0)$   is zero, and we can conclude that the highest degree coefficient $\alpha_0$ should be equal to -4 ($\alpha_{0I}=-4, \alpha_{0R}=-4$), yielding

\begin{equation}
\begin{array}{l}
      -4 K^{2}+12K-9=0   
 \implies K=\dfrac{3}{2} \\ 
 \end{array}
\end{equation}
\begin{center}
\begin{table*}[htb]
\centering
\caption{Rendez-vous/Intercept responses for maximum initial velocity heading. Conditions: $\color{black} x(0)=x_0,y(0)=0, C_I=0,~{\bf k=0}$ (constant wind). For the rendez-vous $\to$ $\color{black} v_{gxf}=v_{gx0}, v_{gyf}=-v_{gy0}$.
}
\begin{small}
\label{table:draglift2}
\begin{tabular}{ccccc}
\hline 
\hline 
$K$ &  terminal $v_{g}$ & position & velocity & acceleration\tabularnewline
\hline 
~$1,3$  & \begin{tabular}{@{}c@{}}free \\ (1D-Intercept)\end{tabular}  & \begin{tabular}{@{}c@{}}linear ($K=1$) \\ cubic ($K=3$) \end{tabular}  &  \begin{tabular}{@{}c@{}}$v_{gx}$ constant,~ $v_{gy}(t)=0$\\ $v_{gx}(t)$ quadratic,~$v_{gy}(t)=0$\end{tabular}  &  \begin{tabular}{@{}c@{}}zero \\ linear\end{tabular} 
\\[0.5cm]
~$\frac{3}{2}$   & \begin{tabular}{@{}c@{}}$v_{gxf}=v_{gx0}\neq0$ \\ ~~$v_{gyf}=-v_{gy0}\neq0$\\ (2D-Rendez-vous)\end{tabular}    & cubic & \begin{tabular}{@{}c@{}}quadratic,~~|$v_{gy0}|=|v_{gx0}|$ \\ Max initial vel. heading, ~|$\theta_{MAX}|=45 ~deg$\end{tabular}  & linear 
\\[0.5cm]
~$\frac{3}{2}$    & \begin{tabular}{@{}c@{}}free \\  (2D-Intercept)\end{tabular}   & cubic  & \begin{tabular}{@{}c@{}} quadratic, ~|$v_{gy0}|=\dfrac{|v_{gx0}|}{\sqrt{3}}$ \\ Max initial vel. heading, ~|$\theta_{MAX}|=30 ~deg$\end{tabular}  & linear 
\\[0.5cm]
$\color{black} 1,3$   & \begin{tabular}{@{}c@{}}$v_{gxf}=v_{gx0}\neq 0$ \\ $ \color{black} v_{gyf}=0$ \\ (1D-Rendez-vous) \end{tabular}  & cubic & quadratic, with ~$\theta_{MAX}=0 ~deg  \to  v_{gy0}=0$  & linear
\tabularnewline
\hline
\hline
\end{tabular}
\end{small}
\end{table*}
\end{center}

 \begin{figure}[hbt]
\noindent \centering{}
\ifOneCol
    \includegraphics[scale=0.63]{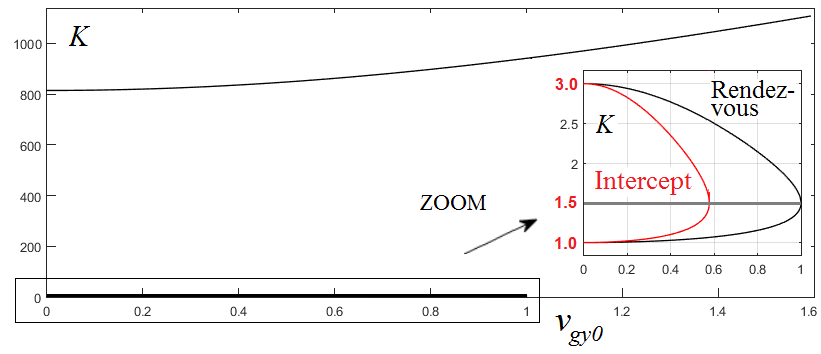}
\else
    \includegraphics[width=\columnwidth]{figs/roots_intercept2z1.png}
\fi
\caption{{Polynomial roots (in multiples of reference time) as a function of $v_{gy0}$, with $C_I=1e^{-6},x_0=3,v_{gx0}=-1.$}}
\end{figure}


Thus, the corresponding critical orthogonal velocity and the maximum initial velocity heading for each case are:
\ifOneCol
    \begin{eqnarray}
     \mbox{Rendez-vous}~~~~~~
    |v_{{gy0}_{MAX}}|=& |v_{gx0}|,~~~~|\theta_{MAX}|=45~ deg \\ 
      \mbox{Intercept}~~~~~~
    |v_{{gy0}_{MAX}}|= & \dfrac{|v_{gx0}|}{\sqrt{3}},~~~~|\theta_{MAX}|=30 ~deg
    \end{eqnarray}
\else
    \begin{eqnarray}
     \mbox{Rendez-vous}~
    |v_{{gy0}_{MAX}}|= |v_{gx0}|,~|\theta_{MAX}|=45~ deg \\ 
      \mbox{Intercept}~
    |v_{{gy0}_{MAX}}|=  \dfrac{|v_{gx0}|}{\sqrt{3}},~|\theta_{MAX}|=30 ~deg
    \end{eqnarray}

\fi

Table 2 summarizes the results.
It is interesting to see that the two different problems above, rendez-vous and intercept, share the same value of critical $K$ (multiple of the reference time), which is $K=\frac{3}{2}$.
The velocity heading limitation for the intercept problem ($30 ~deg$) is in agreement with the ZEM/ZEV results of \cite{Hawking2013,Guo2012Generalz}.
However, there was no previous analysis for the rendez-vous case in the literature, to the best of our knowledge. Note also that the  maximum initial velocity heading for the rendez-vous case will depend on the value assigned to the terminal velocity, as can be seen from the second and fourth lines of Table 2. 

A plot of the $K$-parameter (multiple of the reference time) as function of the orthogonal velocity $v_{gy0}$ for a problem with $x_0=3,v_{gx0}=-1$ is shown in Figure 5 (zoom) for both the rendez-vous and the intercept cases, using $C_I=1e^{-6}$. If instead of considering the simplified quadratic polynomials (with $C_I=0$) we work with the original quartic polynomials in (\ref{eq:poly_compara1b}),(\ref{eq:poly_compara1}), with a very small $C_I=1e^{-6}$, then we have an additional (large) positive root, as shown in Figure 5, and it corresponds to the solution for the minimum-energy problem referred in \cite{Guo2019}. This root will always exist with $C_I \to 0$ because the polynomial is quartic for such cases.
However, there will still be other two local optimal solutions for $t_f$, for a given range of parameter $v_{gy0}$, as indicated in Figure 5.  Note that this figure only shows the three positive real roots of the polynomial, as the fourth one is negative. 

The "bifurcation" in the number of positive real solutions, when $K=1.5$, is clear from this figure. The bifurcation phenomena is the change in the number of multiple roots with the variation of a given coefficient of the polynomial.
The bifurcation was cited in  \cite{Bakolas2014}, for an intercept problem with wind, as "jump solutions" or "manifold discontinuities", although there was no analysis of conditions for the existence of feasible solutions. Also, bifurcation was not discovered before in the literature for the rendez-vous case, to the best of our knowledge.
\textcolor{black}{
In conclusion, we see that the inclusion of $C_I \neq 0$ and/or the presence of a
wind gradient ${\bf k}\neq0$ guarantees at least one positive real root for the polynomial. However, a change in the number of solutions may still occur for a  polynomial with $C_I \neq0$.}

\section{Simulation Results}
\label{eight}

For the case of a constant wind acceleration (section V.B), we consider a rendez-vous problem with the initial conditions
${v}_{gx0}=-1m/s,{v}_{gy0}=0$, ${x}_{0}=30m, y_0=15m$, and terminal conditions ${v}_{gxf}=0,{v}_{gy0}=0$, ${x}_{f}=0$ and $y_f=0$. Supposing a wind speed in the x-direction only, we simulated the optimal guidance control for 
three different $k$ values: $k=-2m/s^2$ (tailwind, increasing with time), $k=0$  (constant wind), and $k=2m/s^2$  (headwind, increasing with time). Two different values of trade-off parameter $C_I$ were considered for each case: $10^{-3}$ and $10$. The corresponding cost-to-go values and optimal travel times are shown at the bottom of Fig. \ref{seven}, together with the resulting optimal paths for each case.
For illustrative purposes, the aircraft heading is also included in the figure, assuming a zero sideslip angle along the trajectory, recalling that the approach is related to a point mass model only. The aircraft shape is plotted at each step of 1.0 sec of simulation.

\begin{figure*}[htb]
\noindent \begin{centering}
\includegraphics[scale=0.36]{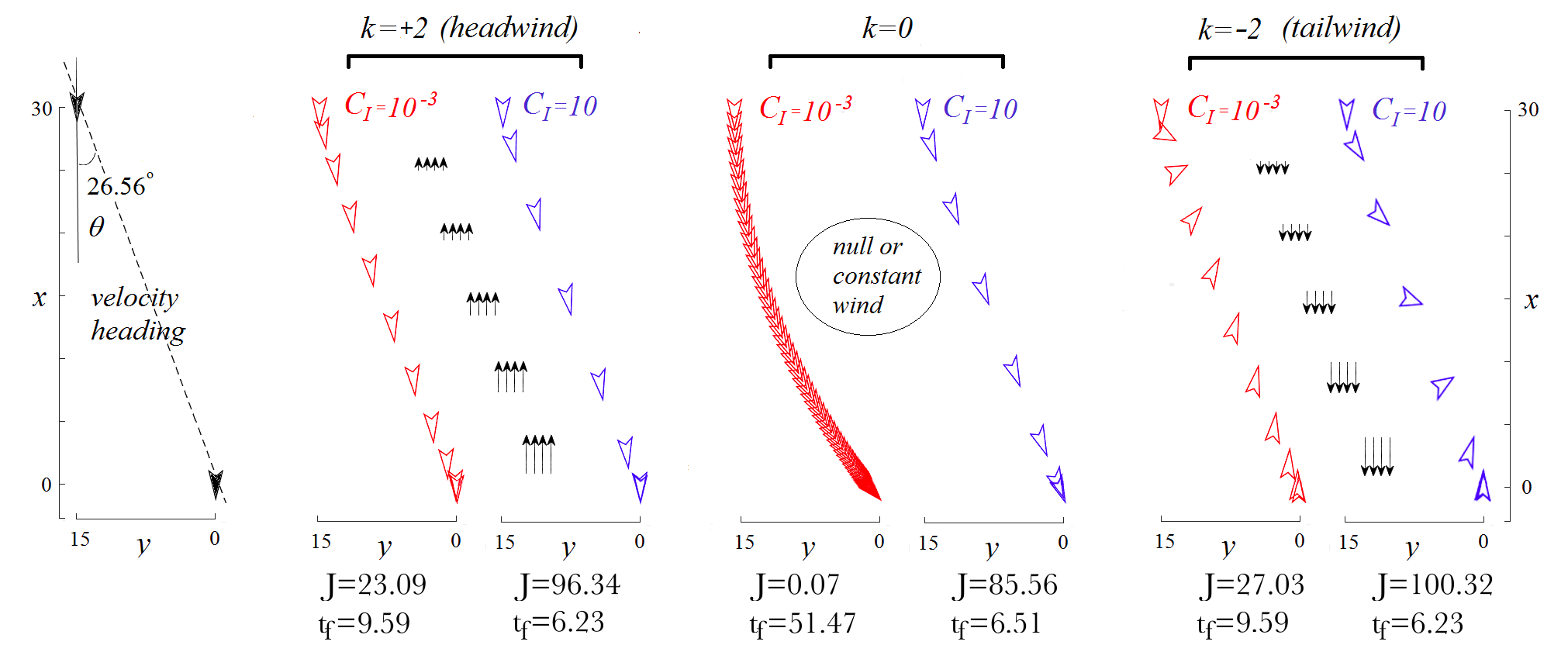}
\par\end{centering}
\noindent \centering{}\caption{Rendez-vous for initial conditions
${v}_{gx0}=-1m/s,{v}_{gy0}=0$, ${x}_{0}=30m$ and $y_0=15m$, and terminal condition ${v}_{gxf}=0,{v}_{gy0}=0$, ${x}_{f}=0$ and $y_f=0$, for 
3 different values of $k$ (wind in the x-direction), and for 2 different values of $C_I$, with corresponding cost-to-go and $t_f$.}
\label{seven}
\end{figure*}

 \begin{figure}[hbt]
\noindent \centering{}
\ifOneCol
    \includegraphics[scale=0.66]{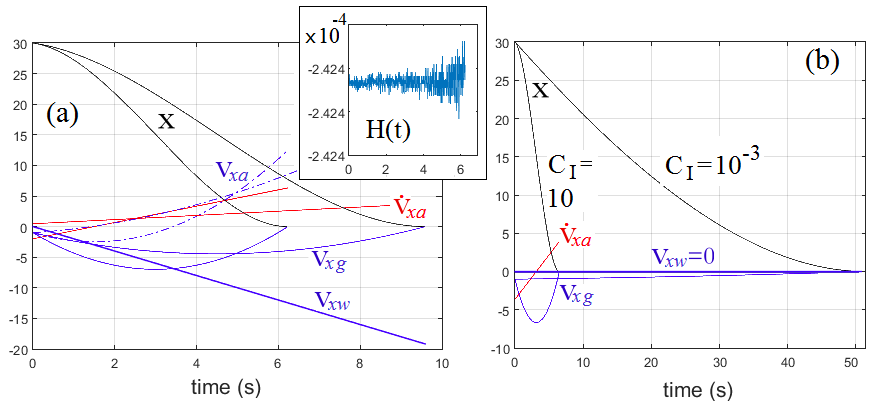}
\else
    \includegraphics[width=0.95\columnwidth]{figs/New_fig04_modBruno1.png}
\fi
\caption{(a) Time responses in $x$-direction for $k=-2m/s^2$ with $C_I=10^{-3},C_I=10$; (b) time responses in $x$-direction for $k=0$  and both $C_I$.}
\label{eight}
\end{figure}

\begin{figure}[htb]
\noindent \begin{centering}
\ifOneCol
    \includegraphics[scale=0.38]{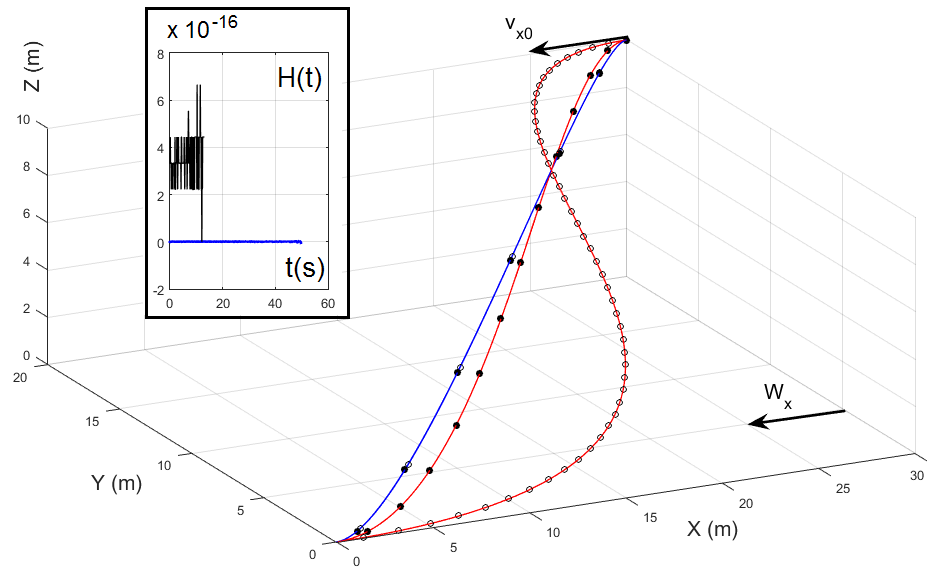}
\else
    \includegraphics[width=0.95\columnwidth]{figs/fig3dd.png}
\fi
\par\end{centering}
\noindent \centering{}\caption{ \textcolor{black}{Rendez-vous in 3D for 
${v}_{gx0}=-1m/s,{v}_{gy0}={v}_{gz0}=0$, ${x}_{0}=30m$,~$y_0=20m$,~$z_0=10m$, and  ${v}_{gxf}=-2m/s,{v}_{gy0}=0,{v}_{gz0}=0$, for 
two different  ${\bf k}$ and $C_I$. Filled balls means a wind rate of ${\bf k}=[-1~ 0~ 0]~m/s^2$, while open balls means zero wind. Red lines are $C_I=10^{-3}$ and blue lines $C_I=10$. In this last case, curves overlap. Hamiltonian $H(t)$ shown at left.}}
\label{nine}
\end{figure}

 \begin{figure}[hbt]
\noindent \centering{}
\ifOneCol
    \includegraphics[scale=0.67]{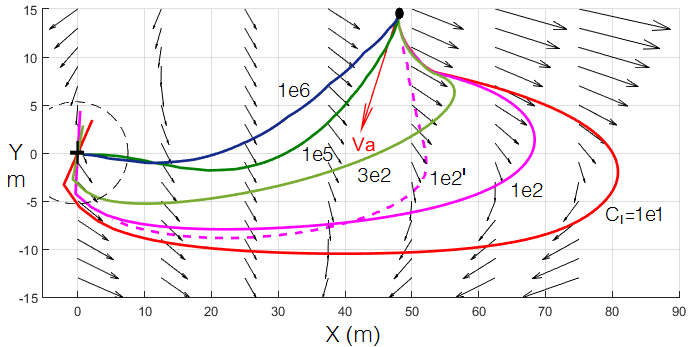}
\else
    \includegraphics[width=0.95\columnwidth]{figs/trajetoria_p3.png}
\fi
{\caption{\textcolor{black}{Optimal trajectories for a piecewise linear wind model, with 5 different $C_I$ values. Dashed line means $C_I=10^2$ case if there was no wind.}}}
\end{figure}

 For the x-direction movement, the corresponding plots of x-position, aircraft velocity (both inertial and relative), wind speed and relative acceleration (control input) are shown in Figure 7 for the cases $k=-2m/s^2$ (a) and $k=0$ (b), and for both values of $C_I$. The corresponding cases can be identified by the respective terminal time at the bottom of Figure 6. 
 \textcolor{black}{At the top of Fig. 7 (a), we plot the time evolution of the Hamiltonian (for $k=-2m/s^2, C_I=10$), while the costate values in this case are ${\bf p}_{r}=[1.33,0.74]^T$ and  ${\bf p}_{v0}=[1.99,2.32]^T$. Recall that as we consider $\dot{\bf w}(t)={\bf k}$, then $H(t)$ is not an explicitly function of time, and thus $H\equiv 0$. This would not be the case if the wind acceleration was a general function of time, or $\dot{\bf w}(t)={\bf f}(t)$.}
 


\textcolor{black}{In Figure 8, we show a simulation case for a 3D rendez-vous with initial conditions
${v}_{gx0}=-1m/s,{v}_{gy0}={v}_{gz0}=0$, ${x}_{0}=30m$,~$y_0=20m$,~$z_0=10m$, and terminal conditions ${v}_{gxf}=-2m/s,{v}_{gyf}=0,{v}_{gzf}=0$, for 
two different values of ${\bf k}$ (no wind and increasing wind toward the target in the x-direction) and two $C_I$ values ($10^{-3}$ and $10$). Note that when $C_I$ is large ($C_I=10$), the presence of the wind makes no difference (as the airspeed is not constrained to be constant), and the final trajectory and travel time are almost the same. However, with a low $C_I$ $(10^{-3}$), when energy saving is important, the results are very different, and the increasing wind toward the target helps in saving energy, as for this case the target velocity is greater than the initial velocity. The plot of the Hamiltonian function $H(t)$ is shown in the same figure, for both wind cases and $C_I=10^{-3}$.
}

\textcolor{black}{
We now present the analysis of a general wind profile case, approximated by a series of piecewise linear time-varying wind speeds, according to the guidelines shown in Section IV.C. This example is similar to those from \cite{2001Airplane},\cite{2008Jardin}, although we consider here variable airspeeds and a trade-off in the performance index. The conditions of the flight are: $(x_0,y_0)=(47.9,14.4)~m$, $(x_f,y_f)=(0,0)~m$. The initial airspeed is $V_a=20~m/s$ with heading angle of $-110~deg$ (red arrow in Fig. 9). The wind speed model is generated by the following spatial distribution: $w_x(x,y)=0.04(x-25)y+4.36$ and $w_y(x,y)=-5.29$. We simulated the trajetory for 6 cases of "base" trade-off parameter $C_I=10,100,300,4000,10^5,10^6$. However, the effective trade-off ($C'_I$) is iteratively adapted following equation (\ref{eq:J_corrected2}), at each $t=0.05~sec$. Within each segment, the wind acceleration is approximated by a constant value, $\dot{\bf w}(t)={\bf k}$. For each of these iteration steps, 10 time step simulations  are done (each time step is $t=0.005~sec$). }

\textcolor{black}{
Some important conclusions can be drawn. Notice that a lower value of $C_I$ tends to make the vehicle follow the wind direction, yielding lower airspeeds, at the cost of longer travel times. The exception is when approaching the target, where large airspeeds occur due to the facing wind. Instead, a larger value of $C_I$ tends to generate shorter paths toward the target, despite the wind speeds, with shorter travel times and higher airspeeds.  It is interesting to note that the adaptive trade-off in a piecewise linear time-varying wind speed is able to guide the vehicle through the wind direction, decreasing its airspeed. Furthermore, the balance between travel time versus control energy (and indirectly with airspeed magnitude) can be achieved by the tuning of the $C_I$ parameter.}

\section{Conclusions}
\label{sec:conclusions}
\textcolor{black}{
This paper proposes an optimal guidance approach for an aerial vehicle navigation using the wind influence.}
The proposed cost function to be minimized involves the weighting of the travel time and the control energy.
Analytical expressions were obtained for both the optimal control input and the corresponding optimal cost in the case of a constant wind acceleration.
The solution is found using Pontryagin’s Minimum Principle and it
was shown to be equivalent to the one obtained with the Zero-Effort-Miss/Zero-Effort-Velocity (ZEM/ZEV) optimal guidance approach.
A fourth order polynomial is proposed whose positive real roots correspond to the optimal travel times.
When  wind is zero, the roots of our polynomial are the same as the solutions presented in the literature for the case of no wind perturbation.
There may exist bifurcation points in the plot of the roots of the optimal polynomial as function of a given initial condition parameter, indicating a change in the number of solutions for the optimal flight time.
We presented an analysis of the number of solutions for the minimum-energy case (trade-off parameter $C_I=0$), which is important also for the classical ZEM/ZEV (without wind).

\section*{Acknowledgments}
The authors acknowledge the funding received from: CNPq DRONI Project (p. 402112/2013-0), Project INCT-SAC - Autonomous Collaborative Systems - (CNPq 465755/2014-3, FAPESP 2014/50851-0), Fapesp BEP (p. 2017/11423-0), Fapesp Auto-VERDE (p. 2018/04905-1), and the Natural Sciences and Engineering Research Council of Canada (NSERC) Discovery's Program.
The first author would like to thank the Concordia Institute of Aerospace Design and Innovation (CIADI) at Concordia University, Montreal, where this work was carried out during a sabbatical leave.

\bibliographystyle{IEEEtran}
\bibliography{Refs}

\end{document}